# Single-Shell NODDI Using Dictionary Learner Estimated Isotropic Volume Fraction

*Abrar Faiyaz[1], Marvin Doyley[1,2,3], Giovanni Schifitto[1,2,4], Jianhui Zhong[2,3,5], Md Nasir Uddin[4]*

[1] Department of Electrical and Computer Engineering, University of Rochester, Rochester, NY, USA
[2] Department of Imaging Sciences, University of Rochester, Rochester, NY, USA
[3] Department of Biomedical Engineering, University of Rochester, Rochester, NY, USA
[4] Department of Neurology, University of Rochester, Rochester, NY, USA
[5] Department of Physics and Astronomy, University of Rochester, Rochester, NY, USA

**Address Correspondence to:**
Md Nasir Uddin, PhD
Department of Neurology
University of Rochester
601 Elmwood Ave
Rochester, NY 14642
Phone: 585-275-5885
Email: nasir_uddin@urmc.rochester.edu

**Abstract Word Count:** 266 words
**Manuscript Word count:** 5,814 words (including in-text citations, but excluding the abstract, acknowledgements, reference list, tables and figure legends)

**Funding information:**
This work was supported by the National Institutes of Health (Grant Numbers: R01 AG054328, and R01MH118020).

**Abbreviations:** dMRI, diffusion MRI; DTI, diffusion tensor imaging; FA, fractional anisotropy; AD, axial diffusivity; RD, radial diffusivity; MD, mean diffusivity; S0, T2 signal with no diffusion weighting; NODDI, neurite orientation dispersion and density imaging; CSF, cerebrospinal fluid; NDI, neurite density index; ODI, orientation dispersion index; $f_{ISO}$, isotropic volume fraction; AMICO, Accelerated Microstructure Imaging via Convex Optimization; MLP, multiple layer perceptron; LTSM, long short-term memory; IHT, iterative hard thresholding; IQT, image quality transfer; T2w, T2-weighted; DLpN, Dictionary-based Learning prior NODDI; DictNet, dictionary based network; GM, grey matter; WM, white matter; IC, intracellular; EC, extracellular; DIPY, diffusion in python; FWI, free water imaging; FW, free water; mFW, multi-shell derived free water; FERNET, Freewater estimatoR using iNtErpolated initialization; ADC, apparent diffusion coefficient; SNR, signal to noise ratio; MEDN, Microstructure Estimation using a Deep Network; PMEDN, advanced MEDN; MESC-Net, Microstructure Estimation with Sparse Coding Network; CSVD, cerebrovascular small vessel disease; SE-EPI, spin-echo echo planar imaging; TR, repetition time; TE, echo time; FOV, field of view.

# ABSTRACT

Neurite orientation dispersion and density imaging (NODDI) enables the assessment of intracellular, extracellular and free water signals from multi-shell diffusion MRI data. It is an insightful approach to characterize brain tissue microstructure. Single-shell reconstruction for NODDI parameters has been discouraged in previous studies caused by failure when fitting, especially for the neurite density index (NDI). Here, we investigated the possibility of creating robust NODDI parameter maps with single-shell data, using the isotropic volume fraction ($f_{ISO}$) as prior. Prior estimation was made independent of the NODDI model constraint using a dictionary learning approach. First, we used a stochastic sparse dictionary-based network (DictNet) in predicting $f_{ISO}$ which is trained with data obtained from *in vivo* and simulated diffusion MRI data. In single-shell cases, the mean diffusivity (MD) and raw T2 signal with no diffusion weighting (S0) was incorporated in the dictionary for the $f_{ISO}$ estimation. Then, the NODDI framework was used with the known $f_{ISO}$ to estimate the NDI and orientation dispersion index (ODI). The $f_{ISO}$ estimated by our model was compared with other $f_{ISO}$ estimators in the simulation. Further, using both synthetic data simulation and human data collected on a 3T scanner, we compared the performance of our dictionary-based learning prior NODDI (DLpN) with the original NODDI for both single-shell and multi-shell data. Our results suggest that DLpN derived NDI and ODI parameters for single-shell protocols are comparable with original multi-shell NODDI, and protocol with b=2000 s/mm$^2$ performs the best (error ~5% in white and grey matter). This may allow NODDI evaluation of studies on single-shell data by multi-shell scanning of two subjects for DictNet $f_{ISO}$ training.

# 1. INTRODUCTION

Neurites are comprised of axons and dendrites and alterations in neurites occur during brain development, normal aging, and in several neurodegenerative disorders [1-4]. Histological analysis of postmortem brain tissue is regarded as the most reliable means of quantifying the neurite morphology and understanding such alterations in neurological diseases. Some limitations of histological assessment include variability in tissue sampling and longitudinal access to tissue. Furthermore, tissue preparation may cause changes in the morphology. Additionally, there are obvious ethical limitations in tissue procurement in living individuals. Conversely, MRI is an excellent modality to provide information about neurites *in vivo,* non-invasively with whole-brain coverage, with faster image acquisition compared to the analysis of histological sections [5].

Diffusion MRI (dMRI) can provide quantitative measures of tissue microstructure by probing molecular water diffusion in the brain [6]. Diffusion tensor imaging (DTI) [7], a widely used dMRI technique, is based on a gaussian water displacement distribution. DTI metrics such as fractional anisotropy (FA) and diffusivities (axial diffusivity AD, radial diffusivity RD, and mean diffusivity MD) are sensitive to alterations in tissue properties, such as cellular density, axon diameter, myelin and water content [6]. However, these metrics are not specific to any particular tissue properties since the DTI model does not account for restricted and hindered water diffusion [8,9].

To address these limitations, advanced multi-shell dMRI technique (i.e., diffusion weighted images acquired with multiple b-values) such as neurite orientation dispersion and density imaging (NODDI) was developed [10]. NODDI, a non-Gaussian model, assumes three types of tissue compartments: intra-neurite, extra-neurite, and cerebrospinal fluid (CSF) that enables the estimation of the neurite density index (NDI), orientation dispersion index (ODI) and isotropic volume fraction ($f_{ISO}$) [aka, extracellular free water (FW)]. NDI measures the neurite density in both axons and dendrites while ODI measures the variability of neurite orientation. These two key indices have been histologically validated by correlating NDI with optical myelin staining and ODI with Golgi staining analysis [11,12]. Several previous studies have demonstrated that NODDI outperforms DTI and provides more insight into the disease processes, normal neurodevelopment and aging [13,14].

However, non-linear fitting in NODDI is computationally intensive and without parallel cluster computing might take long computational times for dMRI data with good resolution (typically ~65 hours) [15]. To reduce the computation time, the problem has been reformatted in a dictionary-based framework by Daducci *et al.* [15] with Accelerated Microstructure Imaging via Convex Optimization (AMICO). The linearization of the non-linear NODDI problem via AMICO resulted in faster convergence when fitting, by four orders of magnitude. Additionally, the calculated parameters were preserved with high accuracy and precision. This also allowed for the sparse reconstruction strategies to be utilized with deep learning approaches to solve the NODDI problem. However, this sparse representation of the dMRI signal in AMICO still requires a large number of diffusion gradients with a multi-shell protocol similar to NODDI, in order to estimate the microstructural parameters.

However, the NODDI model requires a large number of diffusion gradients ( typically ≥90 directions) with multi-shell protocols of at least two b-values (e.g., b=1000 and 2000 s/mm$^2$) [10]. In order to make the protocol clinically feasible, efforts have been made to reduce the number of diffusion gradients using deep learning [16-18]. To the best of our knowledge, no efforts have been made to reduce multi-shell dependency using deep learners. Compared to conventional methods, deep learning approaches are advantageous since the reconstruction of NODDI parameter maps can be very fast once the training is completed, and noise can be reduced via spatial consistency of the dMRI signals. Golkov *et al*. [16] used a multiple layer perceptron (MLP) with three hidden layers to estimate NODDI parameters from multi-shell dMRI data. Subsequently, Ye [17] proposed an optimization-based deep learner [19] which uses *iterative hard thresholding* (IHT) [20] in order to calculate the sparse representation using learned weight vectors that outperforms MLP approach. The network was further improved by incorporating historical information in an adaptive process, by updating the sparse codes that modified the *long short-term memory* (LSTM) units using the IHT process, improving estimation accuracy [18]. However, multi-shell dMRI data was used for both training and test cases for these methods because existing literature highlights that single-shell NDI, ODI and $f_{ISO}$ fitting is ill-posed with NODDI.

Recently, Alexander *et al.* [21] proposed a computational approach called image quality transfer (IQT), illustrating the potential reconstruction of NODDI parameters using diffusion tensors calculated from standard single-shell dMRI data. They built parametric regression trees based on the calculated diffusion tensor features manually to train distributed leaf weight vectors on NODDI generated outputs. In short, the method divided the hyperparameter space based on spatial DTI features and learned discrete weight vectors at the leaf nodes to map microstructural parameters. However, IQT is a DTI feature-based parametric learning where no ground validation is provided for single-shell

NODDI reconstruction. Another study by Edwards *et al.* [22] derived a NODDI-DTI relationship once $f_{ISO}$ was set to a fixed zero using single-shell data in regions with little to no $f_{ISO}$ contamination. These findings reinforced our hypothesis that prior $f_{ISO}$ may be useful in reconstructing NDI and ODI with single-shell data.

We hypothesized that with a NODDI based data driven estimation of $f_{ISO}$, it might be possible to obtain well-posed NDI and ODI maps with single-shell dMRI data. While several techniques are available for $f_{ISO}$ (aka, FW) estimation using free water imaging (FWI)[23-26], these approaches rely on DTI based optimization. Thus, it is expected that the estimated FW does not account for the complex characterization of tissue microstructure similar to NODDI modelling. Recently two studies reported that FW or $f_{ISO}$ could be estimated reliably from single-shell with MD and T2-weighted signal (S0) based initialization technique named Freewater estimatoR using iNtErpolated initialization (FERNET) [26,27], where the single-shell approach is prior driven.

In this work, using both synthetic and *in vivo* data, we formulate a sparse dictionary for $f_{ISO}$, with a learner that replaces the iterations of the IHT technique for sparse dictionary learning with a stochastic layer. Corresponding MD and S0 signals were utilized to build the dictionary for single-shell $f_{ISO}$ estimation. The idea is to address important $f_{ISO}$ priors and complex tissue configurations through the architecture. We then compare the performance of single-shell protocol reconstructions with multi-shell ones generated with Dictionary-based Learning prior NODDI (DLpN) and NODDI for NDI and ODI.

## 2. MATERIALS AND METHODS

In this section, we first briefly review the background of the NODDI tissue model. Then we describe our DLpN fitting. Finally, we present the training and validation strategies using synthetic data simulation and *in vivo* data. The scripts of DLpN fitting are available at https://github.com/abrarfaiyaz/DictNet. This also guides to the sources of relevant simulation data and example HCP data. Further HCP data sources and details have been elaborated in Section 2.4.

Additionally, we used FWI algorithm for single-shell simulation cases which is based on Pasternak's [23] algorithm contained in a singularity container whereas multi-shell FWI [24], [28] was used in multi-shell cases available in the open-source library diffusion in python (DIPY) [29]. PMEDN derived parameters were computed using publicly available tools (https://github.com/PkuClosed/MEDN/)[17].

### 2.1 NODDI tissue model

DTI fitting assigns one apparent diffusion coefficient (ADC) for a single voxel while the NODDI tissue model hypothesizes three different microenvironments, where the ADC in each hypothesized compartment is different. So, for an MR signal in a given voxel, the associated microenvironments are extracellular, intracellular and free-standing water in different configurations. The related diffusion coefficients for three compartments are respectively $d_\parallel$, $d_\perp$ and $d_{ISO}$. The underlying assumption is $d_\parallel$ and $d_{ISO}$ can be assumed with the properties known from *in vivo* cases [10]. Therefore, the NODDI tissue model is stated as follows –

$$A = (1 - f_{ISO})\{\text{NDI} \cdot A_{IC} + (1 - \text{NDI}) \cdot A_{EC}\} + f_{ISO} A_{ISO} \tag{1}$$

For the intracellular (IC) compartment,

$$A_{IC} = \int_{S^2} f(\vec{n}) e^{-bd_{\parallel}(q.n)^2} d\vec{n} \quad (2)$$

where b = b-value, q = b-vector, n = samples of directions on a sphere on which the integration is done. Probability of finding orientation directed along *n* when *μ* and *κ* known,

$$f(\vec{n} \mid \vec{\mu}, \kappa) = M\left(\frac{1}{2}, \frac{3}{2}, \kappa\right)^{-1} e^{\kappa(\vec{\mu}\cdot\vec{n})^2} \quad (3)$$

In the Eq. 3, *M* is defined as Kummer's confluent hypergeometric function.

For the extracellular (EC) compartment,

$$A_{EC} = e^{-bq^T D_{EC}(f,\text{NDI})q} \quad (4)$$

where,

$$D_{EC}(f, \text{NDI}) = \int f(\vec{n}) D_h(\vec{n}, \text{NDI}) d\vec{n} \quad (5)$$

Details for Eqs. 4 and 5 can be found in the previous report [10].

For the isotropic compartment,

$$A_{ISO} = e^{-b.d_{ISO}} \quad (6)$$

where $d_{ISO} = 3.0 \times 10^{-3} \text{mm}^2 s^{-1}$

ODI is dependent on *κ,* known as the concentration parameter associated with the Watson distribution defined in Eq. (3) and calculated as follows

$$\text{ODI} = \frac{2}{\pi} \arctan\left(\frac{1}{\kappa}\right) \quad (7)$$

We use multi-shell protocol with ground-truth parameters to synthesize the diffusion weighted signal based on the defined model and Rician noise with signal to noise ratio (SNR) of 20 dB is added to the synthesized signal.

**2.2. Dictionary-based Learning prior NODDI (DLpN)**

Figure 1 illustrates the schematic of the proposed DLpN network. The DLpN approach separates the estimation of $f_{ISO}$ adopting a dictionary based learning strategy termed as "DictNet" (Section 2.3), and then fits for the other non-linear parameters posed in the NODDI problem with a Rician noise model [10]. In this framework, the NODDI toolbox [10] was modified to use $f_{ISO}$ as a known parameter. Once we have an approximation of the $f_{ISO}$, the Rician loglikelihood framework can be employed to solve the inverse problem of identifying NDI and ODI. Considering the estimated likelihood based on the dMRI signal, the initial parameters are selected from the grid search. Since $f_{ISO}$ is already approximated from the dictionary framework, the grid search complexity is reduced.

We fit the parameters by minimizing the *negative* of Rician loglikelihood defined as follows:

$$L(\text{NDI}, \kappa, \theta, \phi \mid f_{ISO}, d_{\parallel}, d_{ISO}) = -\log \prod_{i=1}^{N} \frac{M_i}{\sigma^2} e^{-\frac{(M_i^2 + A^2)}{2\sigma^2}} I_o\left(\frac{AM_i}{\sigma^2}\right) \quad (8)$$

which is similar to the NODDI problem, except that the initial $f_{ISO}$ prior is estimated with DictNet. In the Eq. (8), $A$ is synthesized and $M$ is the measured signal, NDI is the neurite density index, $\kappa$ term is inversely proportional to ODI (Eq. 7), $f_{ISO}$ is the isotropic volume fraction, and $\sigma$ is the estimated standard deviation from the measured signal per voxel, $\theta$ and $\phi$ are the fiber directions initially estimated from the weighted least squared DTI fitting, then fitted with NDI and $\kappa$ in the NODDI framework.

## 2.3 DictNet

DictNet is a sparse dictionary based learning strategy that has been devised based on previous deep learner models [17,18] to estimate f$_{ISO}$. Typically, with a known sparse dictionary Φ, the coefficients *f* can be assessed by an ℓ1-norm regularized least squares problem

$$\hat{f} = \arg\min_{f \geq 0} \| \Phi f - y \|_2^2 + \lambda' \| f \|_0 \tag{9}$$

where, $\lambda'$ is an adjustable parameter to control the sparsity level of $f$. This was resolved using the IHT algorithm [20] by the following formulation used in Microstructure Estimation using a Deep Network (MEDN) [17]

$$f^{t+1} = H_\lambda(Ld_{\text{in}} + Df^t) \tag{10}$$

where, *t* is the iterative index, *L* and *D* are layers determined by the sparse dictionary Φ, and $H_\lambda(.)$ is a thresholding function with $\lambda > 0$,

$$H_\lambda(x) = \begin{cases} 0, & x < \lambda \\ x, & x \geq \lambda \end{cases} \tag{11}$$

The thresholding function is defined by a parameter $\lambda$ which is related to $\lambda'$, $d_{\text{in}}$ is the normalized diffusion signal cascaded with 3×3×3 spatial data.

    Previously proposed deep architecture models such as MEDN, advanced MEDN (PMEDN) [17] and Microstructure Estimation with Sparse Coding Network (MESC-Net) [18] used 8 iterations of IHT in its original and modified form. However, this redundant iterative

process can be reduced by seeding of a stochastic vector to resemble a generative model. By replacing the iterative scheme with a constant stochastic layer, we can eliminate unnecessary weight vectors thereby saving memory and training time as follows for the estimation of $f_{ISO}$ of a voxel without hindering the performance.

$$f^I = H_\lambda(Ld_{in} + Dd) \tag{12}$$

This is illustrated in Eq. (12), where $d$ is the constant stochastic vector, the basis on which the dictionary is built; Now, to incorporate important $f_{ISO}$ priors such as MD and S0 in the dictionary Φ, the following layers were added, accounting for Eq. (12).

$$f = H_\lambda\big(MD \cdot L_{MD} + S0 \cdot L_{S0} + f^I \cdot L_{n_{ISO}}\big) \tag{13}$$

where, $L_{MD}, L_{S0}, L_{n_{ISO}}$ weighs $MD$, $S0$ and $f^I$ respectively in single-shell $f_{ISO}$ learning. The generated coefficient vector $f$ contributes to a fully connected feedforward network to estimate the $f_{ISO}$. The final contribution was threshold with a thresholding ReLU function similar to PMEDN defined in the Eq. (11). However, Eq. (13) is only valid for single-shell cases; for the multi-shell cases the $f^I$ from Eq. (12) contributes to the fully connected feed forward network for $f_{ISO}$ estimation.

    Our DictNet differs from other conventional deep learners (e.g., PMEDN, MLP) in several key aspects:

1. Conventional deep learners for NODDI focused on reducing only the diffusion gradients without changing multi-shell configuration of the protocol whereas our proposed network mimics the behavior of these conventional deep learners with the seeding of a constant stochastic vector and focus on generating $f_{ISO}$ with single-shell.

2. Seeding of the constant stochastic vector guides the *in vivo* training and we sensitize the model on the simulated data in the process which is not done in previous approaches (PMEDN or MLP).

3. Our model accounts for T2w i.e., non-diffusion weighted signal S0 (inherently collected as b0 images) and MD in training. We provide empirical evidence in the NODDI simulation that using S0 and MD allows the model capable to estimate single-shell $f_{ISO}$ more accurately than other approaches.

4. It requires 8 folds less memory and reduced time compared to conventional approaches as the stochastic vector initialization helps in quick learning.

## *2.4. Subjects and MR acquisition*

### *2.4.1 In vivo Data*

De-identified MRI images from 8 randomly selected subjects from the publicly available Human Connectome Project (HCP) dataset provided by WU-Minn HCP, release-Q3 [30] were used for this study. Written informed consent was collected from all subjects and the study was approved by the institutional review board (https://db.humanconnectome.org/data/projects/HCP_1200). Multi-shell dMRI images were acquired using a 3T MRI scanner (Connectome, Siemens, Erlangen, Germany) using spin-echo echo planar imaging (SE-EPI) sequence: repetition time (TR) = 5520 ms, echo time (TE) = 89.5 ms, field of view (FOV) = 210 ×180, voxel size = 1.25×1.25×1.25 mm$^3$, multiband factor = 3, bandwidth = 1488 Hz/pixel, 90 gradient directions for each shell: b-values = 1000, 2000, and 3000 s/mm$^2$, and 18 non-diffusion weighted images with b = 0 s/mm$^2$, and scan time for each shell was around 9:50 min. Subject data were corrected for bulk motion, susceptibility-induced and eddy current distortions [31]. Further details of scan parameters and study protocols can be found in http://protocols.humanconnectome.org/.

All the protocols used in this work are tabulated in Table 1. In this work, microstructure parameters (NDI, ODI, $f_{ISO}$) computed by NODDI toolbox [10] with Pall (i.e., full set of 270 diffusion gradients) were considered as pseudo-ground-truth.

In addition, we tested our approach to a participant with cerebrovascular small vessel disease (CSVD) from an ongoing study at the University of Rochester. The participant provided written informed consent prior the scans, and the study was approved by the University of Rochester's Research Subject Review Board. Details of the scan parameters are provided in Supplementary materials.

### *2.4.2 Synthetic Data*

In order to evaluate the DLpN framework together with NODDI we utilized known ground-truth tissue microstructures for different protocols and synthesized MR signals, as similar to the approach used in the original NODDI paper [32], with $f_{ISO}$ = 0 as well as *with* additional $f_{ISO}$ cases reported in Table 2. The $f_{ISO}$ is negligible in WM [33], but in order to investigate the underlying effect of GM structures and free water contaminated ROIs, additional $f_{ISO}$ cases of 0.12, 0.4, 0.75, 1 were added in the simulation experiment. The simulation strategy is described previously [10]; We have used the publicly available NODDI toolbox (version 1.01, https://www.nitrc.og/projects/noddi_toolbox) and modified the initialization to fit the simulation needs for different diameters, *a*. To simulate WM and GM tissue configurations, model parameters were set to representative values for both tissue types. The selected set of parameters in simulation is reported in Table 2. The parameter set accounts for 400 different microstructural configurations in 254 uniformly sampled q-space directions termed as mean orientation, $\mu(\theta, \varphi)$. Different mean orientations were used to create synthetic training and test dataset. FSL's "*dtifit*" tool was used to calculate the MD and S0 of the synthesized signals.

## *2.5. DictNet: Train, Validation and Test*

Initially, the network is trained with the simulated data. The synthetic simulated dMRI data (Section 2.4.2) with Rician noise accounts for a greater number of tissue configurations than the tissue configurations expected from a single brain. The initial state of the network is obtained by minimizing error on the simulated data for the stochastic vector *d*.

$$N_0 = \Upsilon_S \left[ \min_d N(d, S) \right] \quad \textbf{(14)}$$

where $S$ is the synthetic training set, $N$ is the current network state, $N_0$ is the updated state obtained at the end of current epoch, and $\Upsilon_S$ denotes training based on minimized network state by $d$ with data $S$. Seeding is performed 10 times (chosen empirically), and the state that minimizes the error gets updated as the current state of the network and further trained with the synthetic training set. If the $d$ vector cannot result in a minimized state for the current epoch, training takes place with the current network state. The training and validation loss per epoch for $f_{ISO}$ with DictNet and PMEDN are presented in Supplementary Figure S1, and details are provided in the Supplementary Material.

*In vivo* training starts from final $N_0$. With 8 randomly selected subjects, we used HCP dMRI data from 2 subjects as the training set, and the remaining subjects were used for testing. The number of training subjects was chosen based on results reported in

Supplementary Figure S2. We found that if the number of training subjects is greater than three, the network is more prone to overfitting. Overfitting after training with 3 subjects is seemingly because the training was performed voxel-wise and for any brain in general the number of possible tissue configurations are limited. So, the increment of training subjects indicates the increment of additional sets of similar data samples, which the network is already familiar with. For two subjects, we had masked brain voxels (each around 145×174×145 samples minus the background) for training where 15% of the data was used for cross-validation in each epoch, with a maximum of 10 epochs. In the train/test phase, the dMRI signal was normalized with mean non-diffusion weighted b0 image and used as input for the network.

For *in vivo* training, the pseudo-ground-truth microstructure parameters (NDI, ODI, $f_{ISO}$) were computed by NODDI [10] using the full set of 270 diffusion gradients. For DLpN reconstruction, we focus on retrieving the $f_{ISO}$ prior, thus the $f_{ISO}$ prior is trained and estimated from the DictNet. The other pseudo-ground-truth parameters - NDI and ODI were used to evaluate DLpN NDI and ODI reconstructions based on the percentage differences and maximization of the objective function. In order to make the dictionary stable for single-shell cases, MD and S0 was obtained for both synthetic data and *in vivo* using a single-shell protocol and, were used in building $f_{ISO}$ dictionary.

## *2.6 Data analysis*

Harvard-Oxford (subcortical GM) and JHU-ICBM (WM) atlases were used to calculate regional averages in standard space (1mm) in pre-defined ROIs. Pearson correlation tests were used to test the associations between two variables. A p-value of < 0.05 was considered statistically significant.

All data processing and analyses were performed using Python (v2.7.16), Keras (v2.0.5), MATLAB 2019a (*MathWorks Inc., Natick, MA, USA*), FSL (v6.0.0) and ANTs (v2.1.0).

## 3. RESULTS

### *3.1 Synthetic data simulation*

We reconstructed the NODDI parameters from synthesized dMRI signals with protocols defined in Table 1 by initializing the model with known $f_{ISO}$ introduced with random 0 to 5% error. The simulation results obtained with $f_{ISO}=0$ were the same as in the original NODDI paper [10] (not shown). However, with additional $f_{ISO}$ cases we found different result and our precursory investigation with synthetic data simulation supports that NDI and ODI can be reconstructed reliably from single-shell dMRI imaging data if $f_{ISO}$ is used as prior. The estimation results for NODDI with $f_{ISO}=0$ and additional $f_{ISO}$ cases are described below.

### *3.1.1 NDI and ODI estimation with and without $f_{ISO}$ prior*

Figure 2 illustrates the NDI reconstruction for different protocols with DLpN and original NODDI fittings. However, using the synthetic data generated with additional $f_{ISO}$ plausible cases, we show that NODDI fitting resulted in NDI deviation (upward bias) from the ground-truth with multi-shell protocols for NDI ground-truths of ≤0.4 (Figure 2). We also found the downward bias with high variance for single-shell protocols reconstructed with NODDI model except at the ground-truth NDI=0.2. In contrast, we observed that $f_{ISO}$ prior reconstruction in DLpN can result in NDIs with markedly improved accuracy and precision in both single- and multi-shell cases. This illustrates that independent estimation of $f_{ISO}$ may lead to a better estimation of NDI. However, some deviations from the ground-truth values were observed in the case of protocol P2 and P3 at lower NDI. This could probably be because measurements at high b-values may not support tortuosity constraints (in the GM or lower NDI region) posed by the NODDI model [34] or lower SNR for higher b-values.

Figure 3 illustrates that ODI reconstruction with proposed DLpN and NODDI fittings for different protocols and ground-truth values. We found similar trends in ODI for both DLpN and NODDI fittings, however DLpN had a lower variance for all protocols. Consistent with previous NODDI reports [10], bias and variance are low for the ground-truth values of ODI less than 0.5. Variability of ODI for higher ground-truth values is related to the ODI itself. Physically, orientation distributions corresponding to large ODIs (e.g., 0.5 to 1) are not very different from one another and the high variance reflects the lack of difference due to its inherent mathematical definition. More details can be found in the original NODDI report [10]. It should be noted that a higher ODI corresponds to highly dispersed neurites, mainly residing in GM and has been previously shown to improve with the number of gradient directions used without any dependency on the number of shells [10].

### 3.1.2 Comparison of $f_{ISO}$ estimation with NODDI, DictNet, PMEDN, and FWI

Figure 4 shows simulation results of the $f_{ISO}$ (aka, FW) estimated with our DictNet and compare the results with NODDI, PMEDN and FWI approaches at 5 different ground-truth values shown for 4 different NDI values. We revealed that DictNet outperform other approaches in both single- and multi-shell cases. PMEDN is the next best for both single- and multi-shell protocols, however, higher deviations are observed in single-shell cases.

In contrast, NODDI and FWI based $f_{ISO}$ show higher variability in most cases.

*3.2 Human brain data*

*3.2.1 NODDI parameters and difference maps*

Figure 5 (A) represents $f_{ISO}$ maps reconstructed using DictNet and NODDI for single- and multi-shell protocols. The $f_{ISO}$ difference maps between the pseudo-ground-truth (i.e., NODDI$_{Pall}$) and DictNet along with NODDI with both single- and multi-shell reconstructions are shown in Figure 5 (B). From the difference maps, it is evident that DictNet outperforms NODDI for single-shell protocols (P1, P2, P3) as well as for two-shell protocols (P12, P13, P23) with respect to the pseudo-ground-truth. Similarly, The NDI and ODI maps were shown for DLpN and NODDI fittings with different protocols in the human brain respectively in Figure 6 (A) and Figure 7 (A); and corresponding difference maps are shown in Figures 6 (B) and Figure 7 (B).

*3.2.2 ROI analysis*

The performance of single-shell $f_{ISO}$ with DictNet, and NDI and ODI with DLpN have been assessed for both WM and GM by means of the percentage error (mean and standard deviation) as presented in Figure 8. The percentage errors were shown for all the parameters and protocols with respect to the pseudo-ground-truth NODDI$_{Pall}$ parameters and reported with the mean of the ROIs. The single-shell protocols (P1 and P2) had around 5% error with small variance in estimating NDI and ODI [Figure 8 (B and C)]. Comparison of the ROI mean values for 6 test subjects with DLpN and NODDI are reported in the Supplementary Figures S3, S4 and S5 (for $f_{ISO}$, NDI and ODI respectively). Pearson correlation analysis for JHU-ICBM WM and HO GM ROIs of the test subjects showed strong correlations between NODDI$_{Pall}$ and DLpN (Supplementary Figure S6).

Correlation coefficients for NDI with the single-shell DLpN with NODDI$_{Pall}$ is highlighted as follows: DLpN$_{P1}$ (r$^2$=0.875), DLpN$_{P2}$ (r$^2$=0.927), DLpN$_{P3}$ (r$^2$=0.944).The DLpN based ODI maps also retained strong correlations with NODDI$_{Pall}$ (r$^2$>0.930) and were found to be similar to those reported previously [10], for both single and multi-shell cases.

### 3.2.3 Comparison of Objective function histograms

In addition, we compared the objective function histograms of all the voxels from the HCP test subjects for both DLpN and NODDI derived metrics and shown an example case representative of all other cases. (Figure 9). Histograms for all metrics were calculated with the objective function defined as the natural log distribution of the Rician loglikelihood i.e. ($\tilde{L} = \ln|L/N|$), where $L$ is the negative Rician loglikelihood described in Eq. (8), $N$ is the number of diffusion gradients for normalizing the objective function.

By the definition of the objective function, the distribution with higher number of voxels that maximizes $\tilde{L}$, can be said to have identified better optimized set of NDI, ODI and f$_{ISO}$. We found that DLpN$_{P1}$ has the highest distribution with maximum $\tilde{L}$ and then DLpN$_{P2}$ and NODDI$_{Pall}$ and the poorly optimized case is found for DLpN$_{P3}$. Region **A** in Figure 9 is defined as the set of bins where NODDI results in voxels with lower objective values but DLpN bins are empty.

### 3.2.4 Multi-shell DTI based FW as prior

In Supplementary Figure S7, we used well-established DTI based multi-shell fitted free water (mFW) (i.e., f$_{ISO}$) and used it as a prior for one of the HCP test subjects. Reconstructed NDI maps from mFW based approach did not reflect with multi-shell fitted (NODDI$_{Pall}$) NDI, because the mFW map is optimized leveraging bi-tensor model, which doesn't account for complex microstructures as NODDI does. The result complements the necessity of NODDI data driven prior estimation of $f_{ISO}$ (or FW).

### *3.2.5 Pathology Investigation*

We applied our DLpN approach to a study participant with white matter hyperintensities characterized by a Fazekas score [35] of 3, and lesion volume of 1.77 cm$^3$). Supplementary Figure S8 presents the NDI maps (coronal view) computed with DLpN (single shell b=2000 s/mm$^2$) and original NODDI (two-shells b=1000, 2000 s/mm$^2$), and corresponding T2 FLAIR images showing lesions. Lesions in both NDI maps are clearly visible. Further details are available in the Supplementary materials and Figure S8.

## 4. DISCUSSION

In this work, we demonstrate that NODDI parameter maps such as NDI and ODI can be reconstructed from single-shell dMRI data using a dictionary learner estimated $f_{ISO}$ as a prior. In order to generate $f_{ISO}$, we devise a network, that takes advantage of the IHT strategy used in recent studies [17,18]. We propose a non-iterative scheme of IHT where a constant stochastic layer determines the learning of dictionary coefficients by the spatial-angular sparse dMRI data from the simulated dataset based on the protocol obtained from the *in vivo* data. The generated coefficient vector contributes to a fully connected feed-forward network to estimate the $f_{ISO}$. The network incorporates important determinants of $f_{ISO}$ priors[26] i.e. MD and, T2w signal S0 that facilitates single-shell estimation of $f_{ISO}$ with NODDI based data driven learning. The overall procedure drastically reduces memory requirements and training time compared to the deep learning approaches (used to reduce gradient directions in NODDI)[17,18], and preserves the estimation accuracy and precision in estimating $f_{ISO}$ compared to the ground-truth data.

Using both simulation and *in vivo* data experiments, we evaluated the feasibility of our proposed DLpN approach for single-shell NODDI parameter mapping and compared the results with the multi-shell NODDI$_{Pall}$. The results from both experiments indicated that single-shell NDI and ODI reconstructions are possible with good accuracy in WM and GM ROIs. Our simulation results revealed that DictNet generated single- and multi-shell $f_{ISO}$ values were stable whereas, on the same dataset, NODDI showed ill-posed behavior in $f_{ISO}$ estimation in several cases likely due to NODDI model constraint. That is, in the NODDI fitting, higher $f_{ISO}$ regions have shown to be explained with higher ODI values, that influences the $f_{ISO}$ contribution as well as NDI. Calculating the $f_{ISO}$ from the dictionary instead of grid search shows that the loglikelihood objective function can be better optimized, and also allows for single-shell fitting reliably in the process. DLpN based NDI and ODI reconstructions for single-shell protocols are very close to the ground-truth values and consistently outperformed the original NODDI fittings.

In addition, comparison of the simulation of DictNet derived $f_{ISO}$ with bi-tensor model-based FW (both for single- and multi-shell), deep learner based PMEDN and NODDI in different protocols show that DictNet derived $f_{ISO}$ estimation is stable with single-shell P1 and P2 protocols, compared to that of other approaches (Figure 4). It is because these approaches do not incorporate either S0, MD, or dictionary accounted for simulated complex tissue microstructure.

Using the *in vivo* HCP data, we found that DLpN based single-shell NDI values had around 5% difference compared to the pseudo-ground-truth NODDI$_{Pall}$ in WM and GM for single-shell protocols (i.e., P1, P2 and P3). Interestingly, overall NDI and ODI results suggest that, P2 had a minimum error for both (~ 5%), even when it showed to have a higher difference with our estimated $f_{ISO}$ prior (~30%) but our $f_{ISO}$ prior retained strong correlations ($r^2 > 0.85$) with the same pseudo-ground-truth $NODDI_{Pall}$ (Figure 8 and Supplementary Figure S6). This means that $DictNet_{P2}$ identified a different (mean-shifted) set of $f_{ISO}$ based on the built dictionary which retained the similar solution for NDI and ODI with $DLpN_{P2}$ similar to $NODDI_{Pall}$. The comparison of objective function histograms between DLpN and NODDI for the example test subject highlights this point (Figure 9). In addition, we show that a set of voxels in $NODDI_{Pall}$ optimization has lower objective values (region **A**), where the goal for all the approach was to maximize this objective function. $DLpN_{P2}$ derived NDI, ODI and $f_{ISO}$ parameters resulted in a very similar objective function histogram when compared to $NODDI_{Pall}$, except region **A** where NODDI performed poorly. As $DLpN_{P2}$ shows to reconstruct similar NDI and ODI maps as $NODDI_{Pall}$ (difference ~5%) and multi-shell fitting noise gets reduced in the process. Therefore, $DLpN_{P2}$ is recommended to create single-shell NDI and ODI maps given training protocol as Pall.

Possible reasons for biases in single-shell P1 and P3 include the following: firstly, DLpN ODIs obtained in P1 and P3 are relatively different from $NODDI_{Pall}$ observed from the simulation (also reported in previous study [10]) and resemble P2 or the middle protocol used in the Pall. Secondly, $DictNet_{P3}$ derived $f_{ISO}$ estimation was not stable in simulation, due to the fact that higher b-value images have lower SNR. This also explains why the $DLpN_{P3}$ objective histogram is seen to be left skewed compared to $NODDI_{Pall}$, suggesting there were a large number of voxels that were not well optimized in $DLpN_{P3}$. Based on our simulation results, $DictNet_{P1}$ and $DictNet_{P2}$ derived $f_{ISO}$ values were shown to be stable in simulation. But for the *in vivo* experiment, P2 performed best in generating parameter maps close to the pseudo-ground-truth (i.e., NODDI$_{Pall}$ derived NDI and ODI maps). In case of P1, we saw P1 based ODI estimation was strongly correlated but different (mean-shifted) compared to Pall based ODI. Yet interestingly, P1 based optimization has shown to yield better objective values than NODDI$_{Pall}$. But if we focus on reconstructing NODDI$_{Pall}$ equivalent NDI and ODI as they are histologically validated [36], P2 based DLpN is the approach to take with Pall based training data. Nevertheless, the high correlation of DLpN$_{P1}$ maps with NODDI$_{Pall}$ suggests the possibility of histological correlation to hold valid for DLpN$_{P1}$ as well, and a subject for future study.

So, to compare DLpN derived NDI and ODI with any other NODDI based studies, it is recommended to use the middle protocol, i.e. in our case, P2 when DLpN is trained with Pall (comprising of P1, P2 and P3). We hypothesize observing from the training pattern that to enable multi-shell equivalent reconstruction on P1, we will need to change the training data protocol. That is, we will need to acquire training data with a b-value lower than P1 and another shell with b-value higher than P1. This will be further investigated on relevant datasets and should potentially allow the clinical cases with P1 to perform NODDI investigation.

As an exploratory analysis, we investigated a pathological case (details in the Supplementary materials) with WM hyperintensities which shows DLpN$_{P2}$ derived NDI, is coherent with the FLAIR image and 2-shell NODDI in identifying the WM hyperintensities and was able to provide expected diminished NDI contrast at the lesion ROIs.

Major advantages of the DLpN approach are follows: first, NDI and ODI maps can be further improved with DLpN if independent f$_{ISO}$ or FW can be better estimated by leveraging phantom based $f_{ISO}$ studies in future with single- or multi-shell cases. Second, the use of a single-shell protocol would reduce the scan time by more than 50% compared to the standard NODDI acquisition. Thus, this approach might be useful to obtain NDI and ODI reconstructions using a clinical scanner and in a clinically feasible acquisition time for cases such as stroke, pediatric or emergency subjects with sufficient resolution. Third, the DLpN approach may be applied retrospectively on the existing data collected with a reasonable number of diffusion directions and appropriate b-value (say b=1000 s/mm$^2$). However, prerequisites for using the existing dataset are that two additional subject scans are required with the same scan parameters with 3-shell protocol to train for $f_{ISO}$ with the test case single-shell protocol as the middle protocol of the multi-shell training set.

The limitation of this study is that we did not use bias field corrected S0 maps, for a fair comparison with the NODDI derived parameters. This can be addressed and evaluated in a future study. Secondly, the nature of datasets used in our study is of relatively high resolution. However, as the training is performed voxel-wise, we expect to see similar results on low resolution clinical dMRI datasets. Thirdly, the DLpN processing time is currently 10 to 13 hours whereas the original NODDI model requires 20 to 30 hours to process on our multicore setup (Parallel 24 cores). However, AMICO setup with our single-shell DLpN approach should be able to reduce the NDI and ODI processing time down to <30 minutes. This work could be further extended to account for a reduced number of diffusion gradients, and incorporating quantitative T2 maps with advanced machine learning approaches.

## 5. CONCLUSIONS

We have proposed a dictionary-based learning prior NODDI (DLpN) for estimating NODDI parameter maps from single-shell dMRI data. This study independently estimates the parameter ($f_{ISO}$) responsible for multi-shell requirement of NODDI fitting and shows single-shell NDI and ODI are stable when using $f_{ISO}$ as a prior. The resulting NODDI parameter maps with DLpN are comparable with those reconstructed with multi-shell NODDI and the reconstruction further depends on training and testing protocols for the data-driven strategy. Since compared to other established and available free water estimation strategies we show $f_{ISO}$ prior estimation can be made independent and stable when the learning dictionary is constructed using corresponding S0 and MD, and NODDI data; this data-driven approach may allow retrospective analyses of single-shell dMRI data to measure NDI and ODI with the prerequisite of two additional scans for training DictNet.

## ACKNOWLEDGEMENTS

This work was supported by the National Institutes of Health, NIH (Grant Numbers: R01 AG054328, and R01MH118020). Data were provided in part by the Human Connectome Project, WU-Minn Consortium (NIH Grant Numbers: 1U54MH091657, U01AG052564 and U01AG052564-S1). We would like to thank Alan Finkelstein for his careful linguistic scrutinization, editing and proofreading.

## Declaration of Competing Interest

The authors declare no competing interests.

## Author Contributions

**Abrar Faiyaz:** Conceptualization, Methodology, Software, Formal analysis, Writing - original draft, Writing - review & editing, Validation; Visualization; **Marvin Doyley:** Resources, Writing - Review & Editing; **Giovanni Schifitto:** Resources, Writing - Review & Editing; Funding acquisition; **Jianhui Zhong:** Writing - Review & Editing; **Md Nasir Uddin:** Conceptualization, Methodology, Writing - original draft, Writing - review & editing, Supervision.

# TABLES

**Table 1: The list of imaging protocols used to perform the synthetic data simultation and used for *in vivo* data.**

| Protocols | b-value (number of gradients) |
|---|---|
| P1 | b1000 (90) |
| P2 | b2000 (90) |
| P3 | b3000 (90) |
| P12 | b1000 (90) + b2000 (90) |
| P13 | b1000 (90) + b3000 (90) |
| P23 | b2000 (90) + b3000 (90) |
| Pall | b1000 (90) + b2000 (90) + b3000 (90) |

Note: b-values are in s/mm$^2$; P1, P2, P3 are single-shell; P12, P13, P23 are two-shell; Pall is three-shell protocols. The number in parentheses indicate the number of gradient directions (excluding b=0 images).

**Table 2: The ground-truth parameter values for the synthetic data simulation**

| Parameters | Ground-truth values |
| --- | --- |
| NDI | 0.2, 0.4, 0.6, 0.8 |
| $f_{ISO}$ | 0, 0.12, 0.4, 0.75, 1 |
| *a* (radii) μm | 0.5, 1, 2, 4 |
| $\kappa$ | 0, 0.25, 1, 4, 16 |
| μ (Θ,φ) | 254 uniformly distributed orientations |

Note: NDI, Neurite Density Index; $f_{ISO}$, Isotropic Volume Fraction; *a*, axon radii; $\kappa$, concentration parameter; μ, orientation.

# FIGURE LEGENDS

**Figure 1:** Schematic of the proposed Dictionary-based Learning prior NODDI (DLpN). The stochastic dictionary learning framework (DictNet) estimates $f_{ISO}$ prior and then used with NODDI Rician log-likelihood estimation steps for NDI and $\kappa$ mapping. A) DictNet, B) DLpN.

**Figure 2:** Comparison of NODDI and Dictionary-based Learning prior NODDI (DLpN) for neurite density index (NDI) reconstruction in simulation with defined protocols (P1, P2, P3, P12, P13, P23 and Pall) where NDI ground-truth (GT) values are indicated with dashed lines: A) 0.2, B) 0.4, C) 0.6, and D) 0.8. Synthetic data simulation utilizes the combination of parameters defined in Table 2 to generate diffusion signals with 20 dB Rician noise. Single-shell (P1, P2, P3) NDI reconstruction fails in NODDI, but seems plausible with the DLpN framework using known $f_{ISO}$ prior.

**Figure 3:** Comparison of NODDI and Dictionary-based Learning prior NODDI (DLpN) for orientation dispersion index (ODI) reconstruction in simulation with defined protocols (P1, P2, P3, P12, P13, P23 and Pall) for ODI ($\kappa$) ground-truth (GT) values are reported: A) 0, B) 0.25, C) 1, D) 4, E) 16. Note $\text{ODI} = \frac{2}{\pi}\arctan\left(\frac{1}{\kappa}\right)$. Variability of ODI for higher ground truth at A) and B) is related to the ODI itself. The orientation distributions corresponding to very large ODIs are not very different from one another and the high variance at higher ODI also reflects a lack of any difference.

**Figure 4:** Dictionary-based network (DictNet) estimation of $f_{ISO}$ at ground-truth values A) 0, B) 0.12, C) 0.4, D) 0.75, E) 1 (indicated with a dashed line) on the simulation test set with defined protocols P1, P2, P3, P12, P13, P23 and Pall for different NDI ground-truth values provided in Figure 2. DictNet generated $f_{ISO}$ are compared with other methods such as NODDI, advanced Microstructure Estimation using a Deep Network (PMEDN), and free water imaging (FWI).

**Figure 5**: A) A representative axial view of the isotropic volume fraction ($f_{ISO}$) maps estimated with proposed DictNet and NODDI for different protocols. Note, NODDI$_{Pall}$ $f_{ISO}$ is considered as pseudo-ground-truth $f_{ISO}$ and used as training data for DictNet; B) Difference maps for DictNet $f_{ISO}$ and NODDI $f_{ISO}$ with respect to NODDI$_{Pall}$ $f_{ISO}$ for defined protocols. Intensity scales are shown.

**Figure 6:** A) A representative axial view of the neurite density index (NDI) maps estimated with proposed DLpN and NODDI for different protocols on a human connectome project (HCP) subject. Note, NODDI$_{Pall}$ is considered as pseudo-ground-truth for comparison. B) Difference map for DLpN and NODDI derived NDI maps with respect to NODDI$_{Pall}$ NDI for defined protocols. Intensity scales are shown.

**Figure 7:** A) A representative axial view of the orientation dispersion index (ODI) maps estimated with proposed DLpN and NODDI for different protocols on a human connectome project (HCP) subject. Note NODDI$_{Pall}$ is considered as pseudo-ground-truth

for comparison; B) Difference maps for DLpN and NODDI derived ODI with respect to NODDI$_{Pall}$ derived ODI. Intensity scales are shown.

**Figure 8:** The estimation errors (%) from *in vivo* data in isotropic volume fraction f$_{ISO}$ (A, B), neurite density index NDI (C, D) and orientation dispersion index ODI (E, F) for white matter (WM) and grey matter (GM) using DLpN single-shell P1, P2, P3, and multi-shell protocols P12, P12, P23 and Pall. Errors were calculated with respect to the pseudo-ground-truth (NODDI$_{Pall}$) for six HCP test subjects.

**Figure 9:** Log histograms for the objective values of DLpN P1, P2, P3 and NODDI$_{Pall}$ with a human connectome project (HCP) test subject. Positive direction along the x-axis corresponds to higher objective values. The goal for both NODDI and DLpN is to maximize the objective function. Region **A** is defined as the set of bins where NODDI probability density function has lower objective values than DLpN.

Figure-1

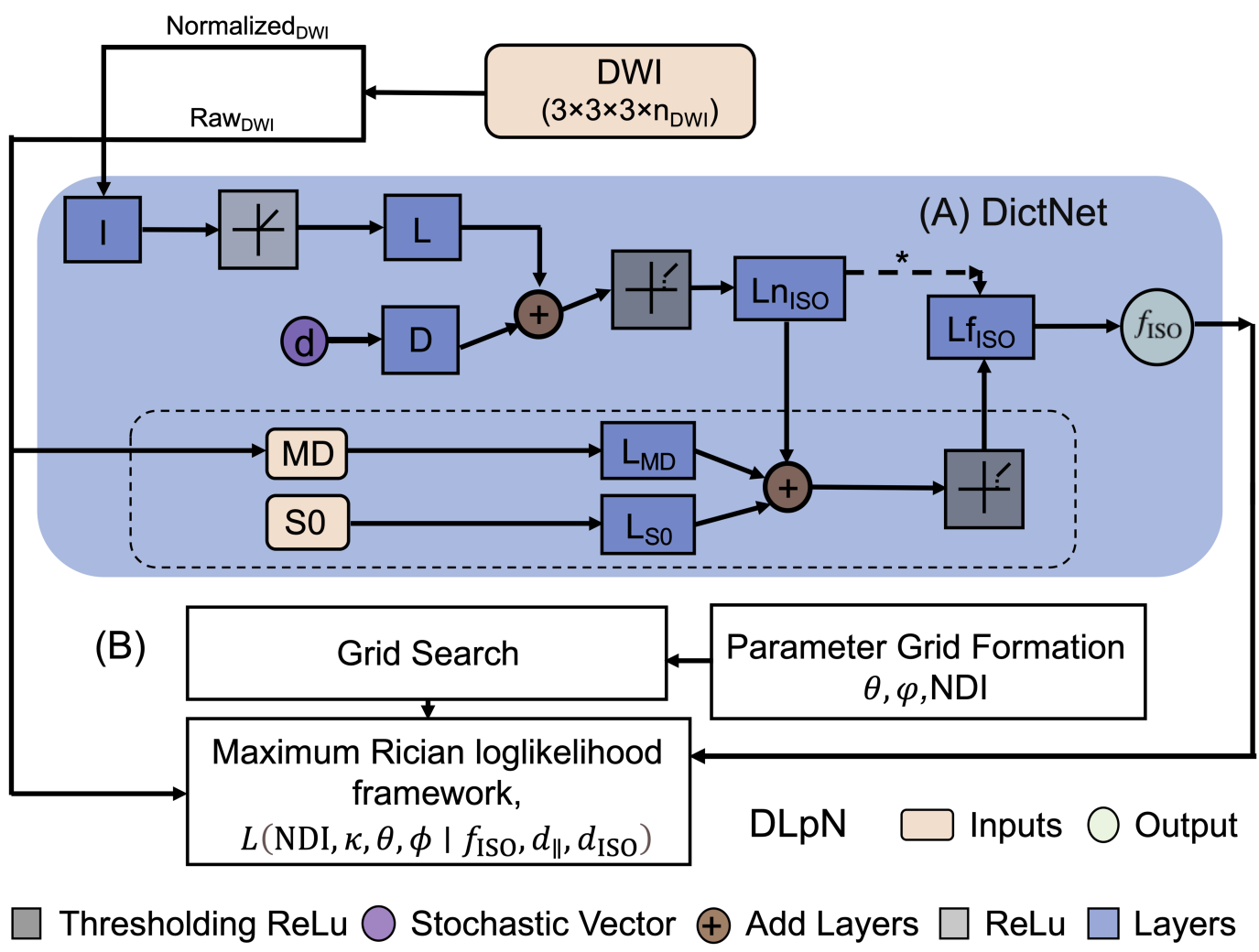

Figure-2

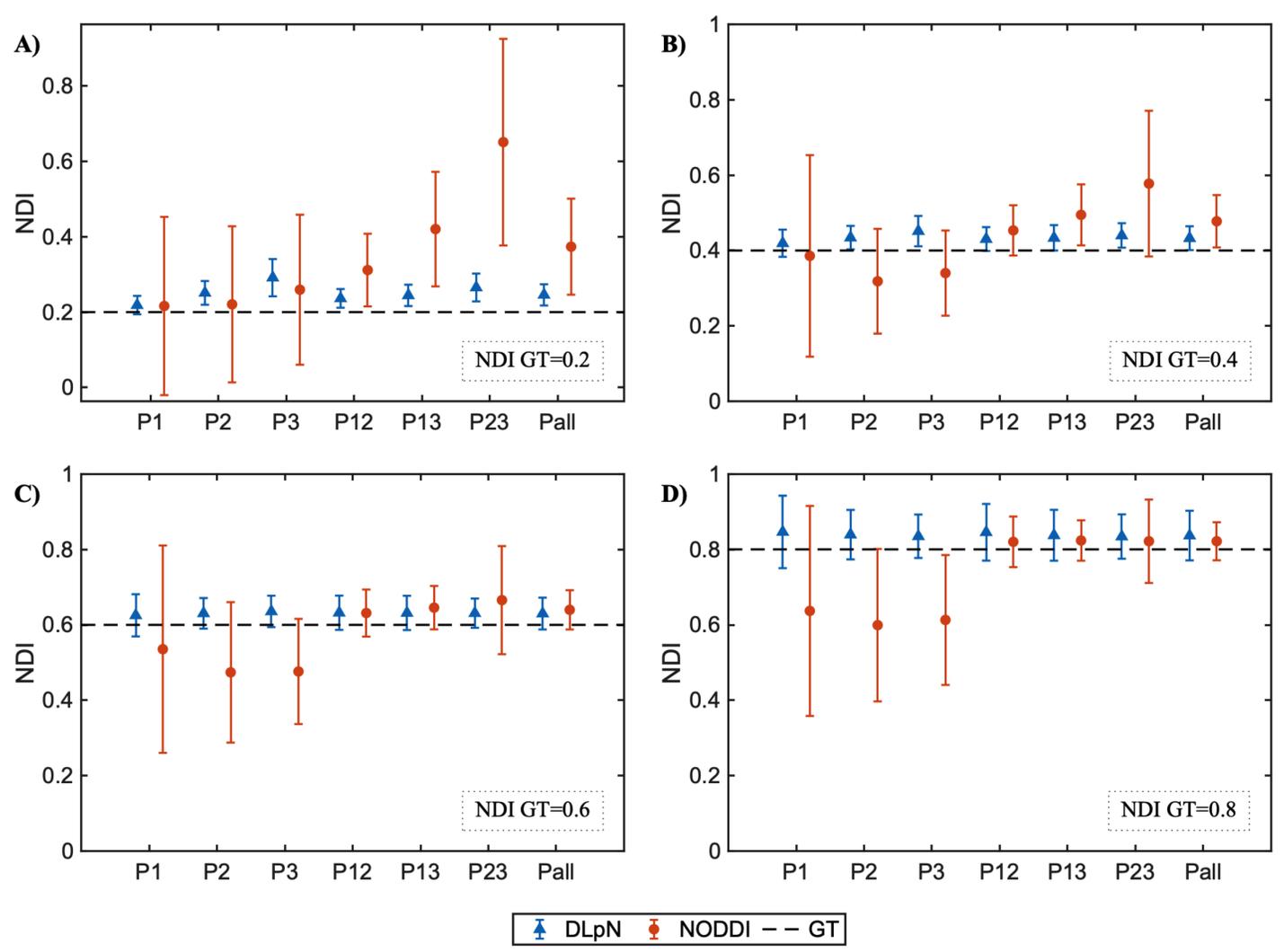

Figure-3

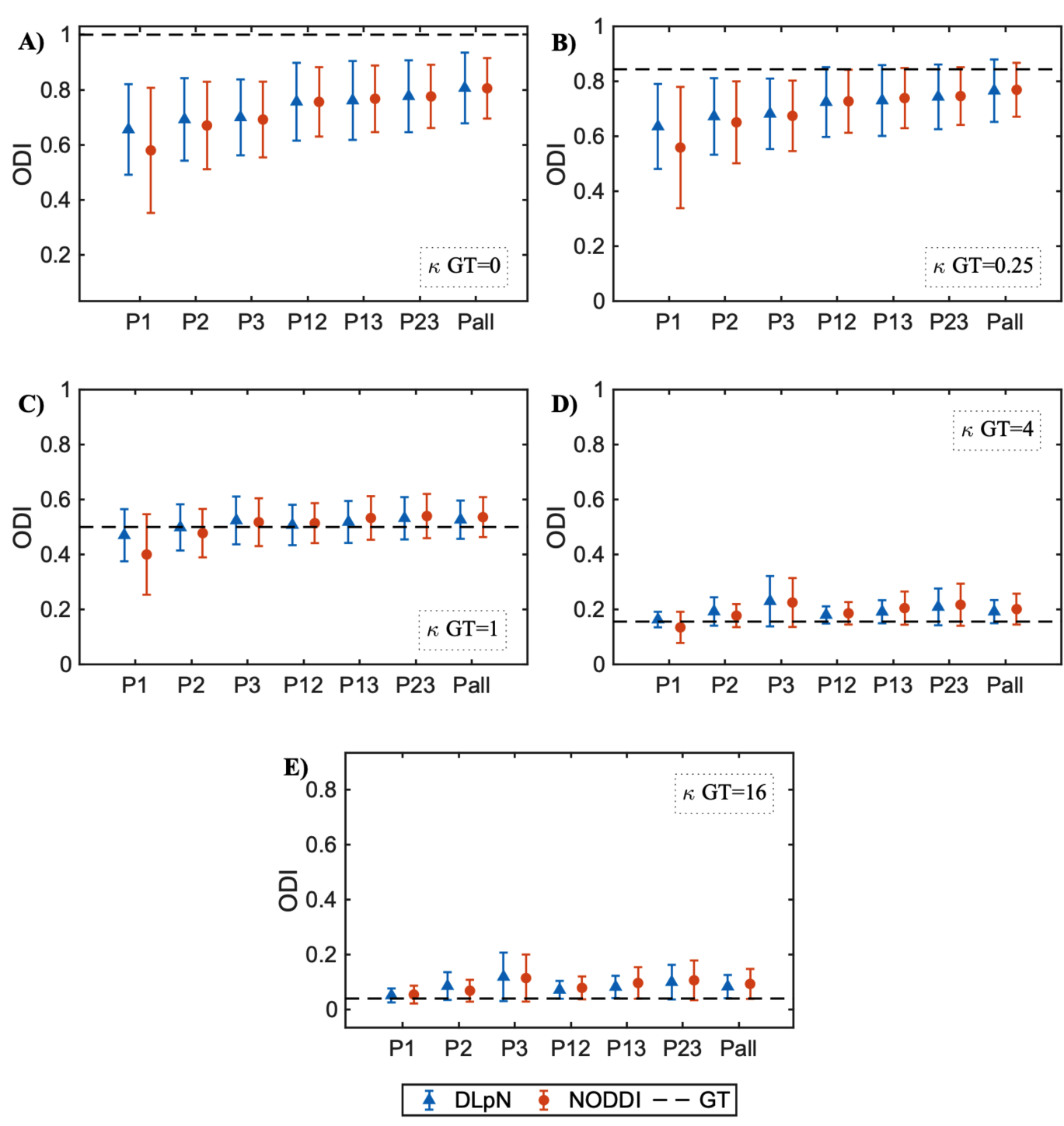

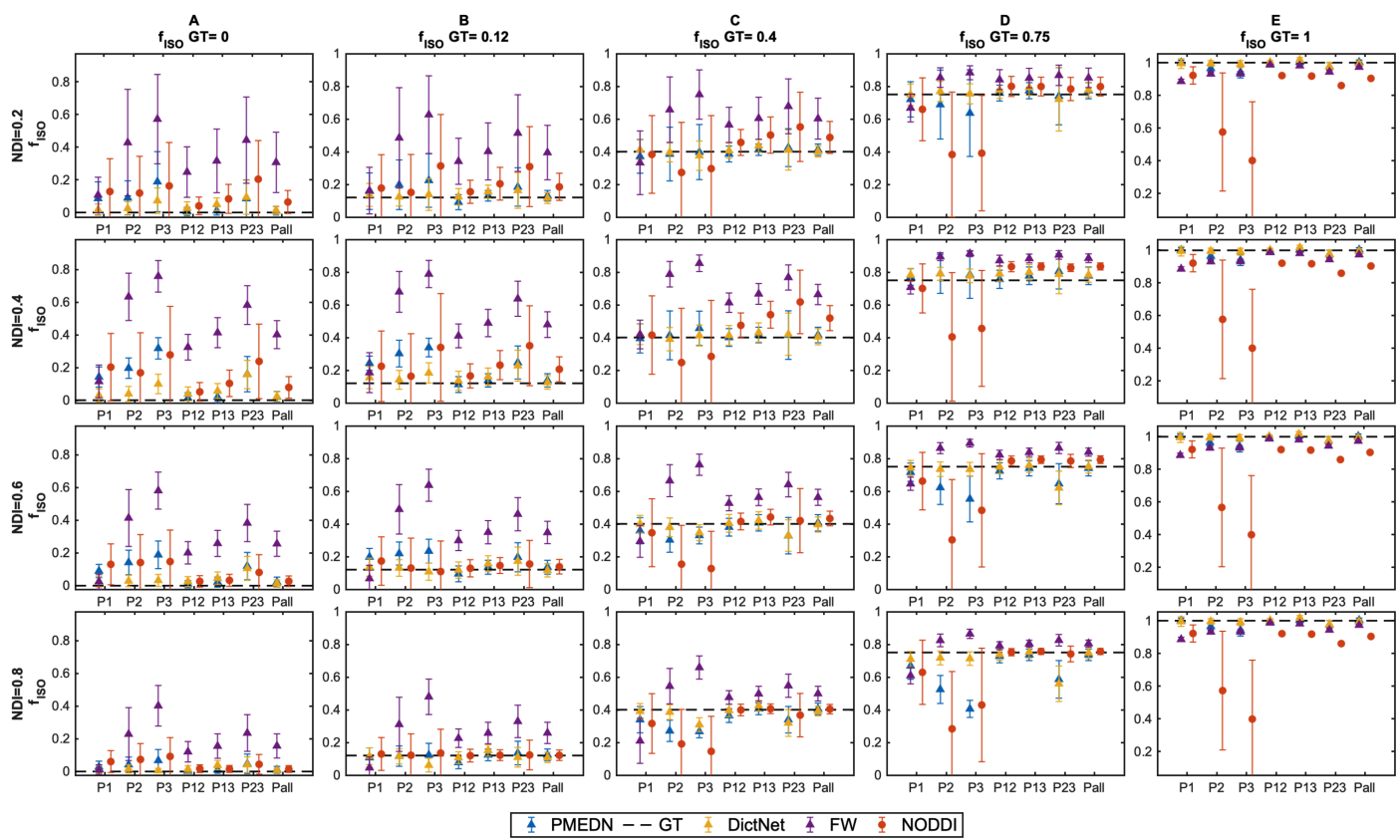

Figure-5

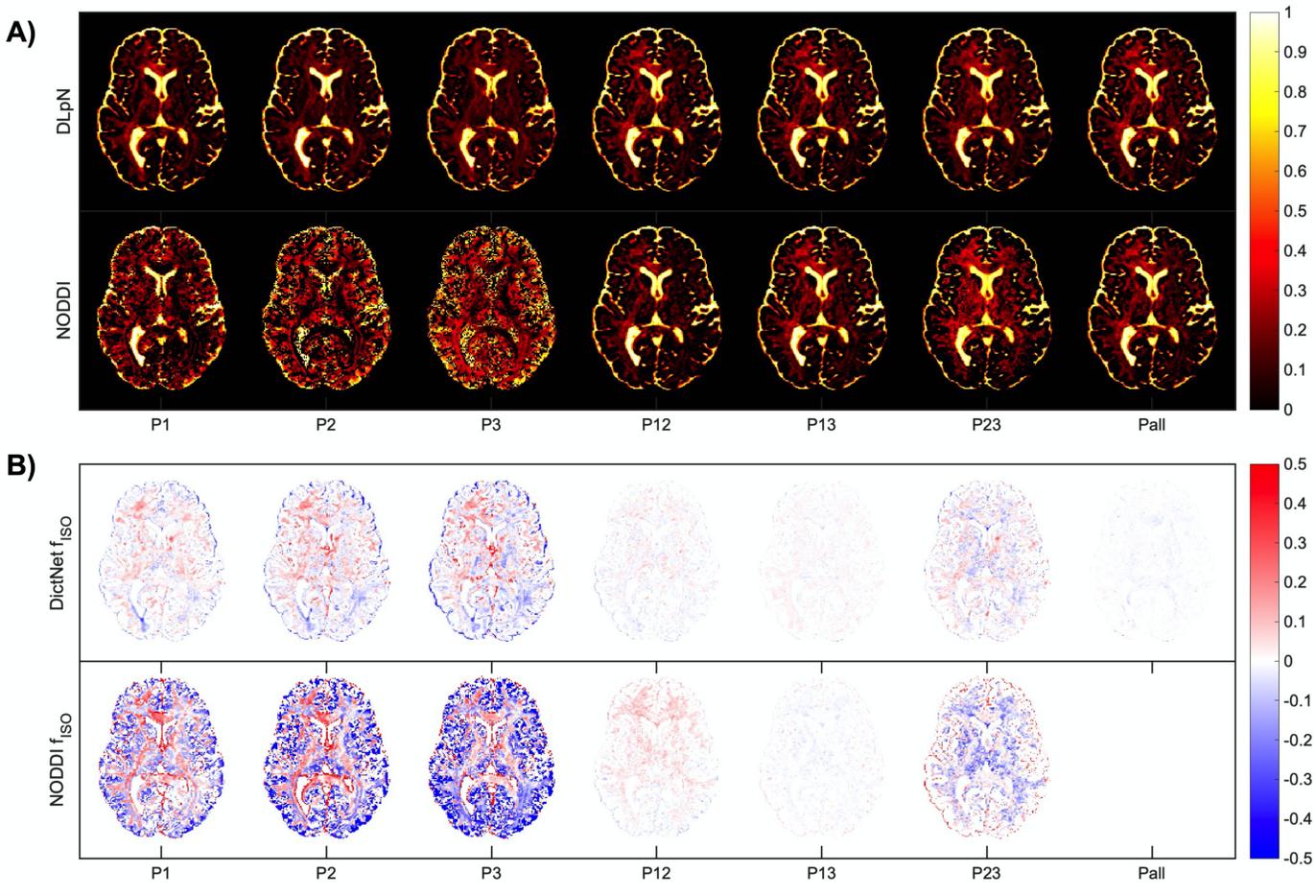

Figure-6

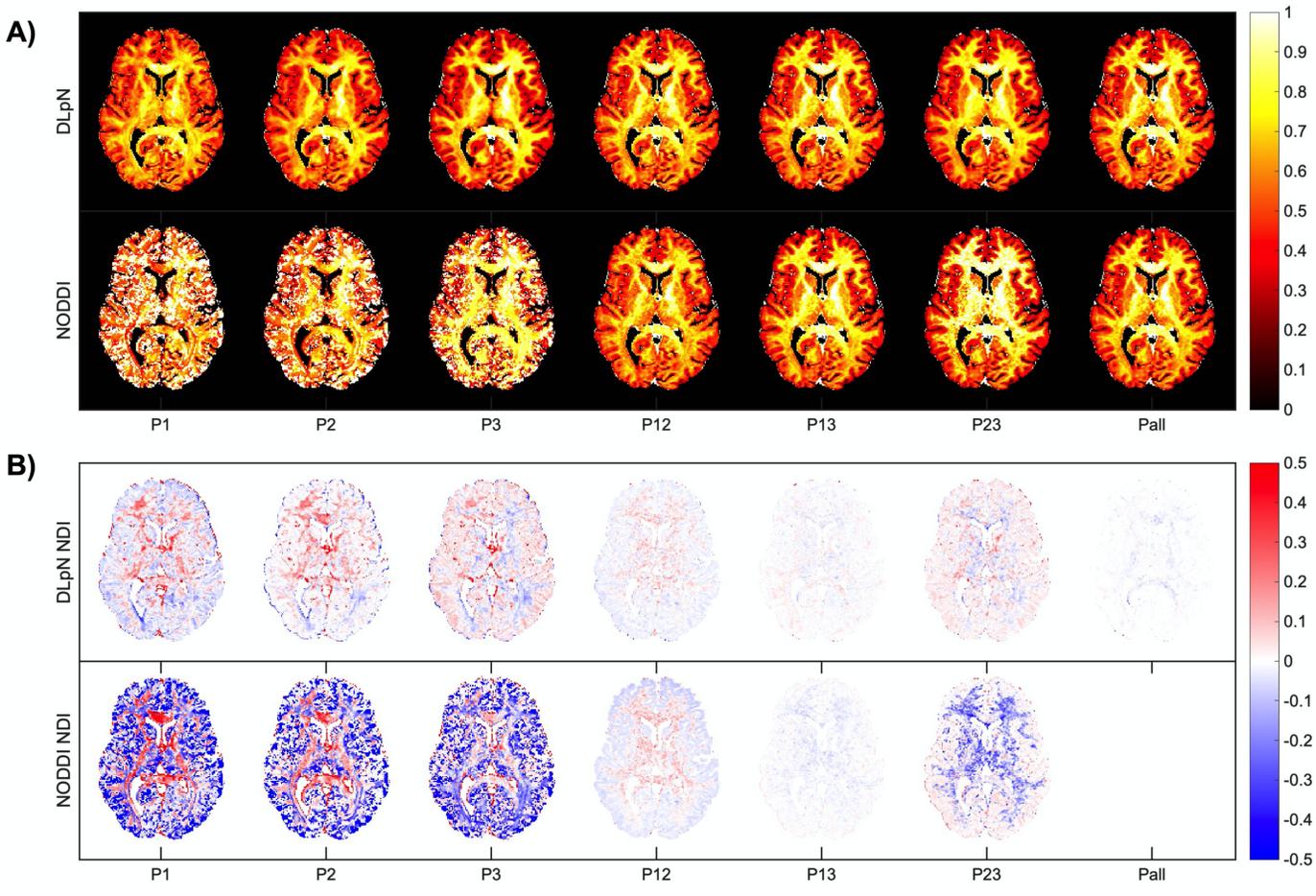

Figure-7

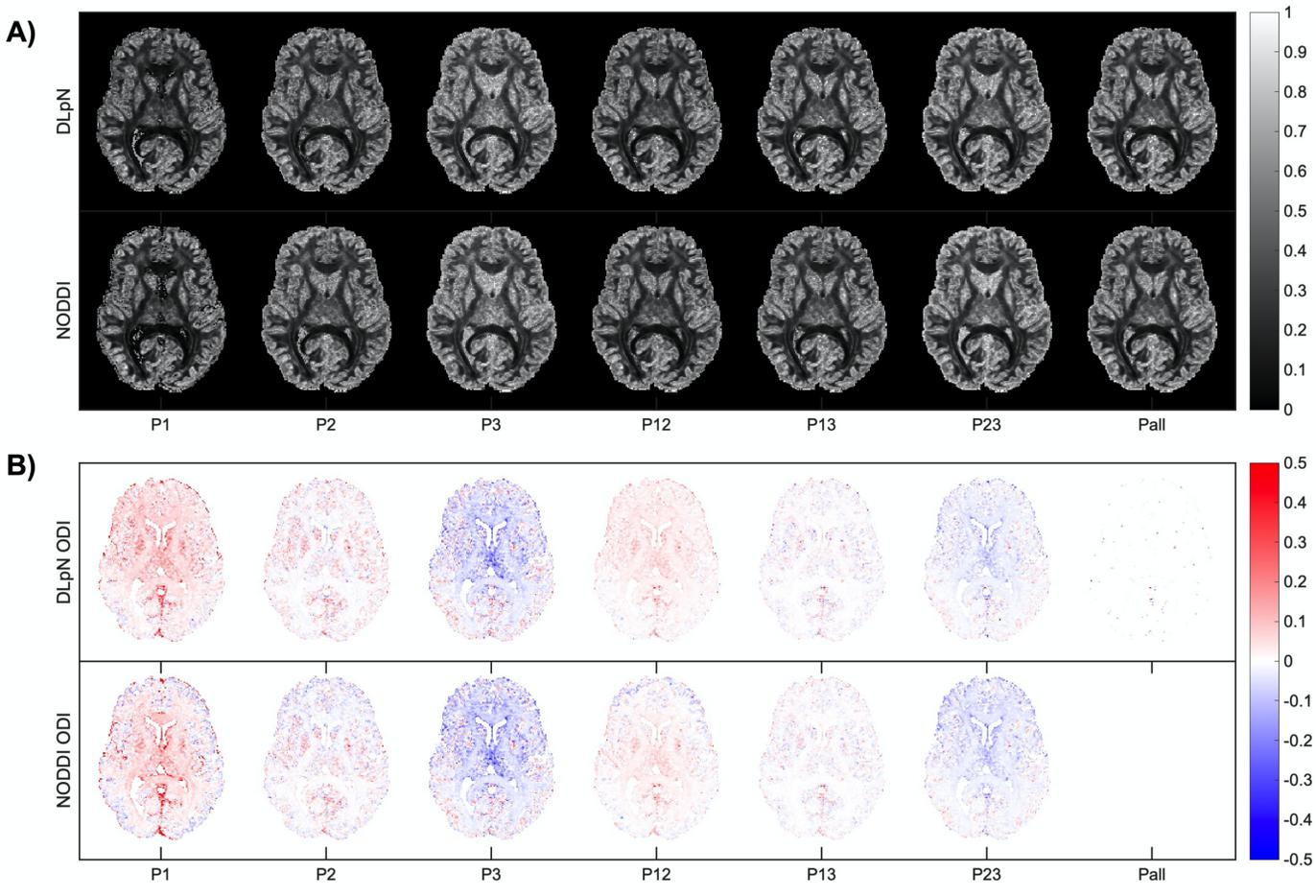

Figure-8

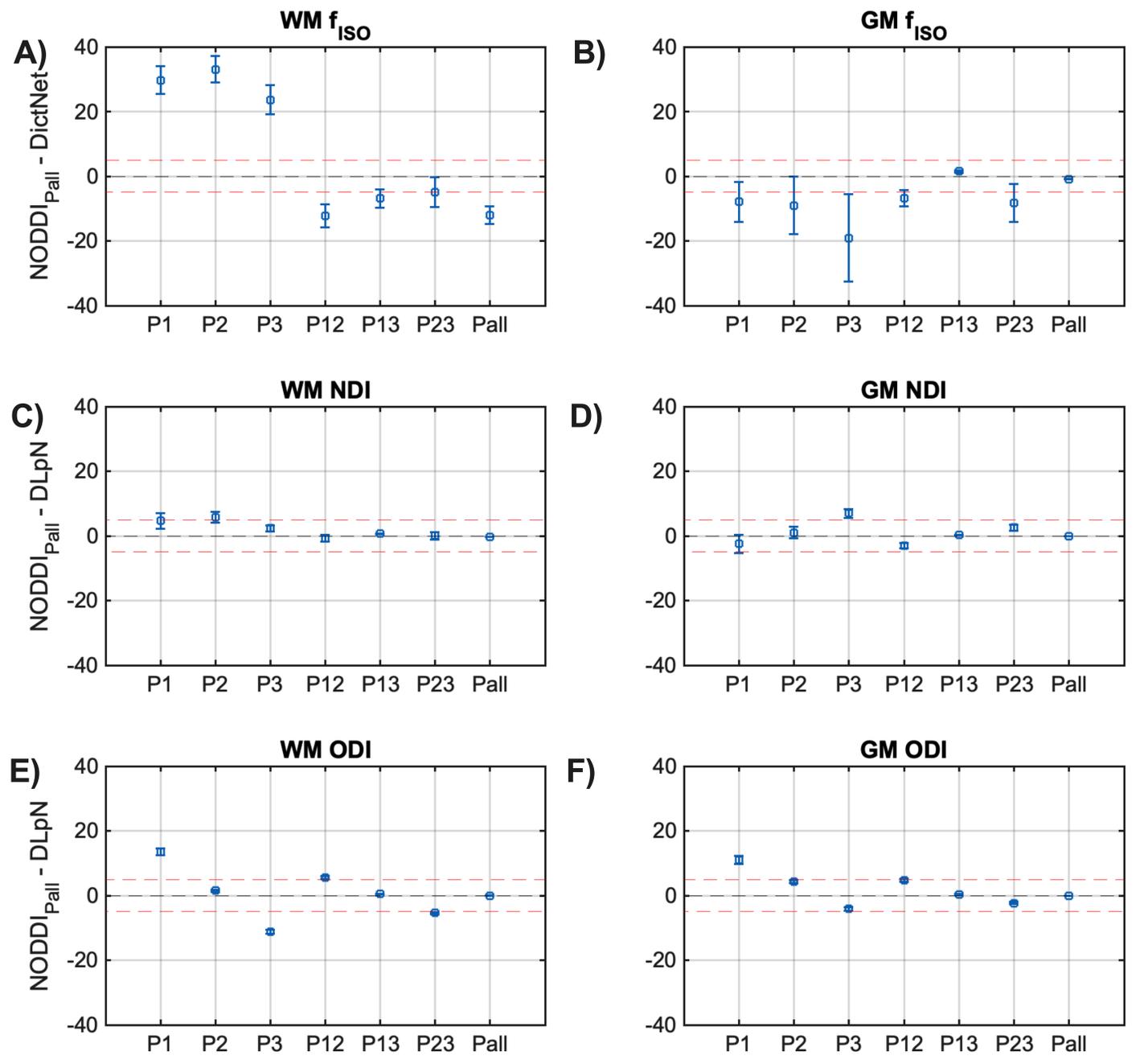

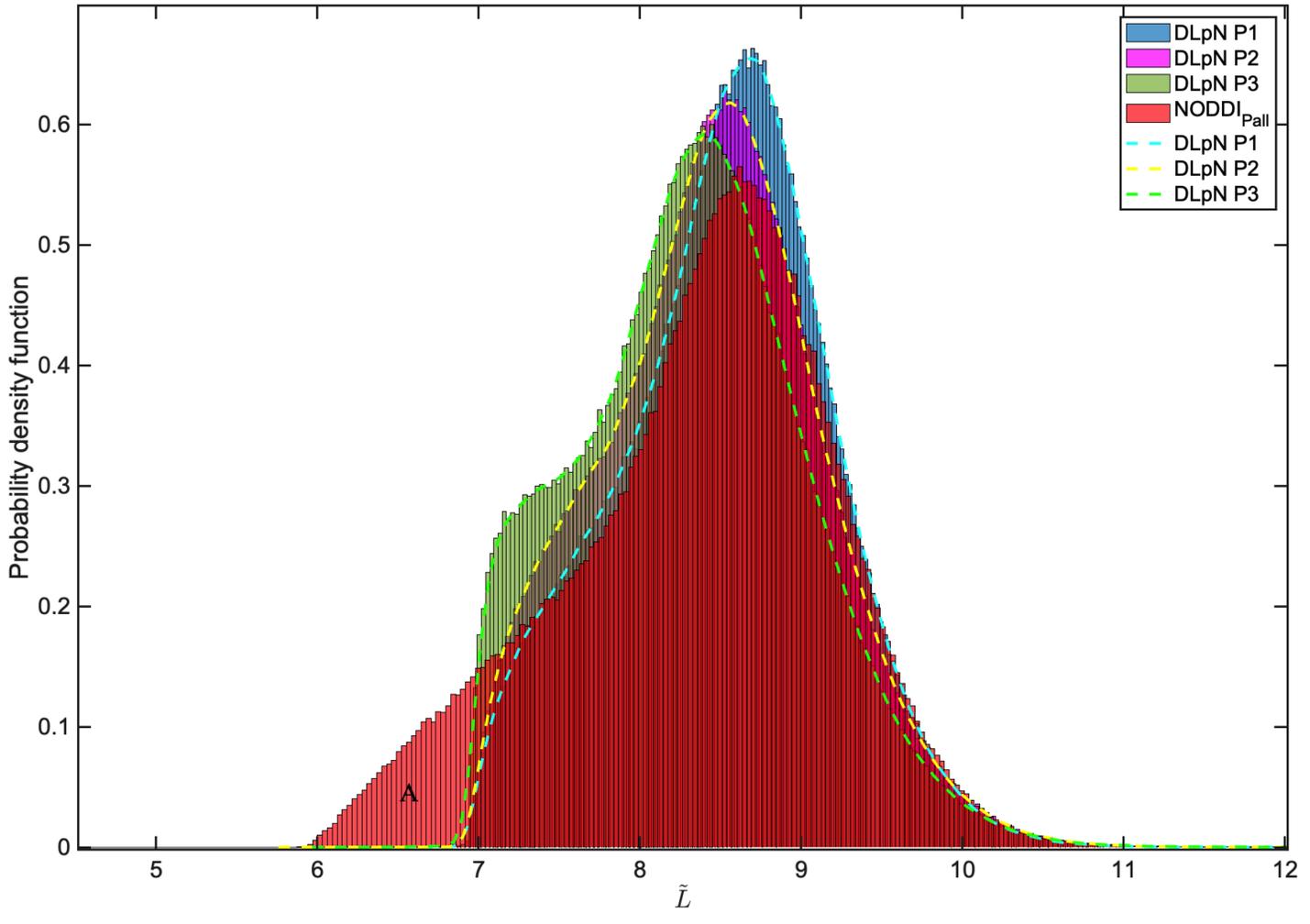

Figure- Graphical Abstract

**HSV NODDI ($f_{ISO}$, ODI, NDI)**

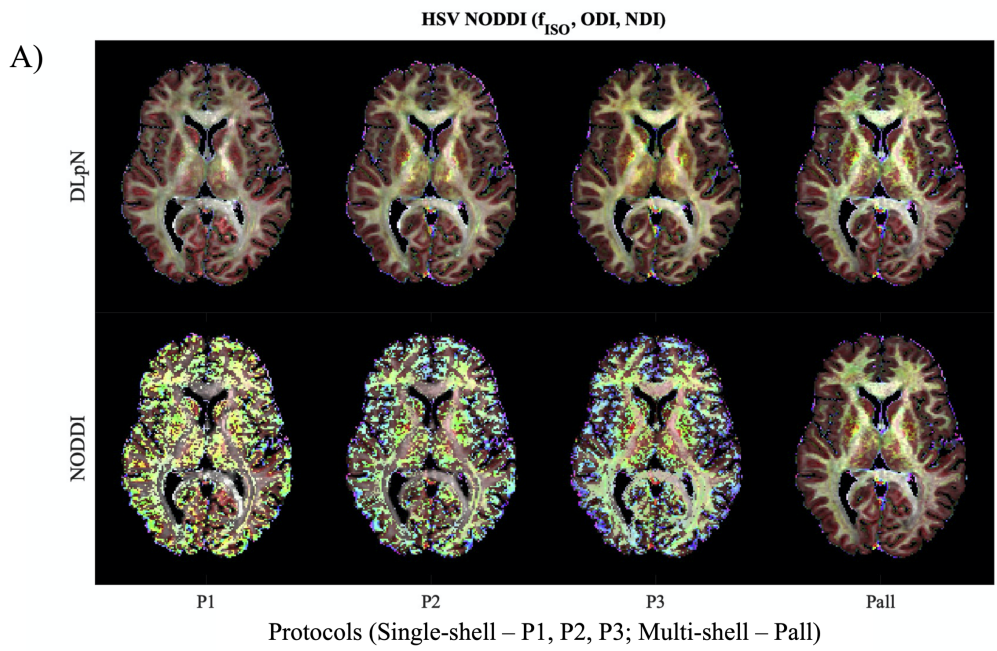

Protocols (Single-shell – P1, P2, P3; Multi-shell – Pall)

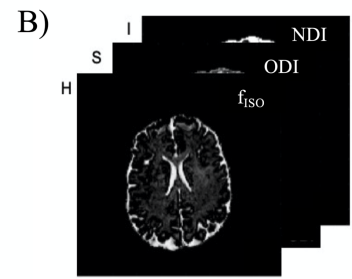

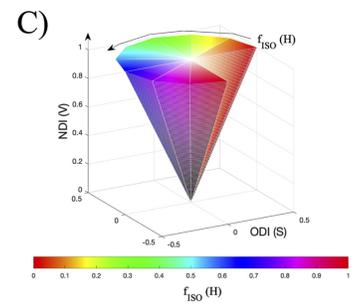

# Supplementary Material

**Title: Single-Shell NODDI Using Dictionary Learner Estimated Isotropic Volume Fraction**

**Supplementary Figure S1. Training and validation**

The training and validation loss per epoch for $f_{ISO}$ with DictNet and extended Microstructure Estimation using a Deep Network (PMEDN): A) single-shell protocol P1, B) multi-shell protocol P12. The results show $f_{ISO}$ estimation error from DictNet can be minimized lower than PMEDN with single-shell and as lower as multi-shell while utilizing less memory and time. Note: P1 = protocol with b=1000 s/mm² and P12= protocol with b=1000, 2000 s/mm².

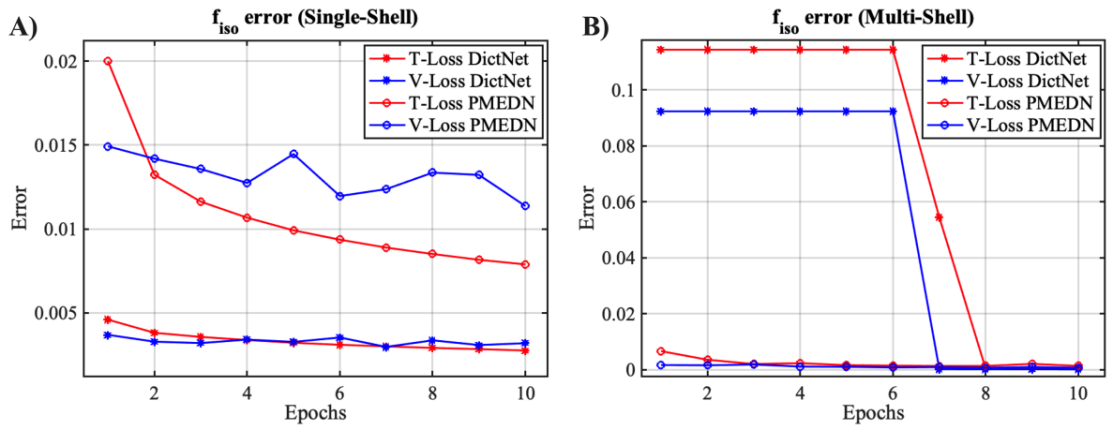

The training and validation loss in the Figure demonstrates the epochs where the minimized state obtained with seeding for DictNet training and validation. Loss epochs for PMEDN is illustrated in the same figure. Figure S1(A) shows a case for DictNet where seeding results in a minimum state at the very beginning of the training for single shell (no other stochastic vector minimizer was obtained in rest cases) and the error then minutely gets reduced with training. Figure S1(B) shows a multi-shell case, where the error was not reducing till the 6th epoch and drops drastically as a stochastic vector minimizer $d$ is obtained at the 7th iterations.

**Supplementary Figure S2. Determining number of training subjects for DictNet**

The sum of mean squared error (MSE) for 4 human test subjects (HCP data) for variable training subjects shown to figure out an appropriate number of subjects required for training. The plot clearly depicts that for voxel-based training strategies, if the number of subjects is increased there are chances of overfitting. DLpN based NDI, $f_{ISO}$ and ODI results indicate increasing trends when training subjects > 3. The network settings were kept the same for all 5 training times. Similar behaviour was observed for PMEDN.

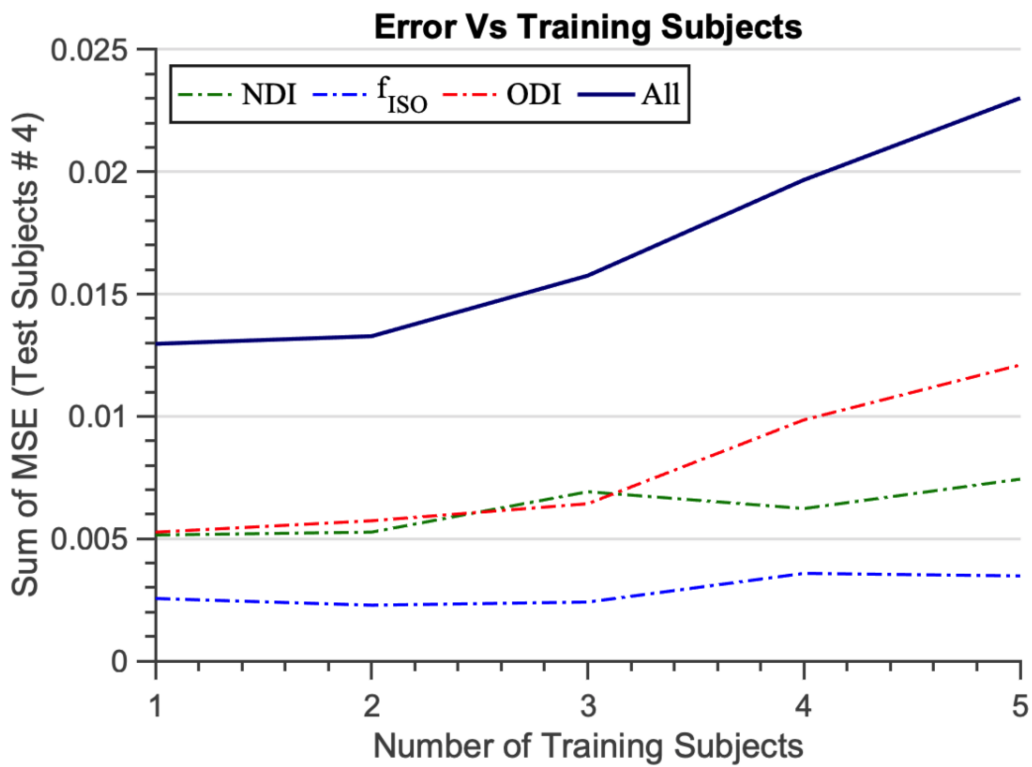

**Supplementary Figure S3. Comparison of $f_{ISO}$ values in brain ROIs**

Comparison of DictNet derived isotropic volume fraction ($f_{ISO}$) with single-shell protocols (P1, P2, P3) and NODDI derived $f_{ISO}$ with multi-shell (NODDI$_{Pall}$) for different white matter (WM) and grey matter (GM) ROIs using John Hopkins University (JHU) WM and WM tract atlas and Harvard-Oxford (HO) GM atlas.

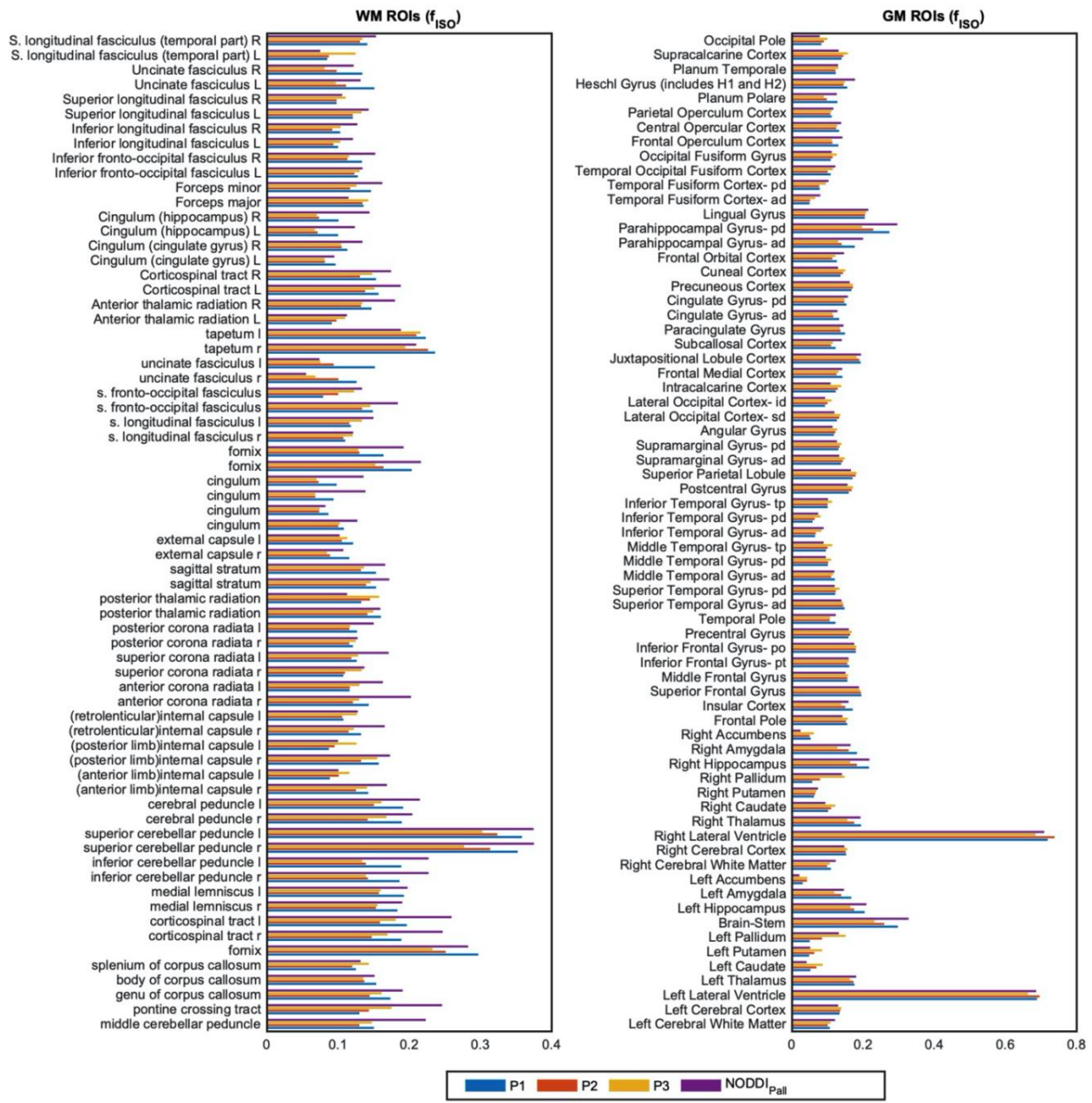

**Supplementary Figure S4. Comparison of NDI values in brain ROIs**

Comparison of neurite density index (NDI) for DLpN single-shell protocols (P1, P2, P3) and multi-shell NODDI$_{Pall}$ for different white matter (WM) and grey matter (GM) ROIs using John Hopkins University (JHU) WM and WM tract atlas and Harvard-Oxford (HO) GM atlas.

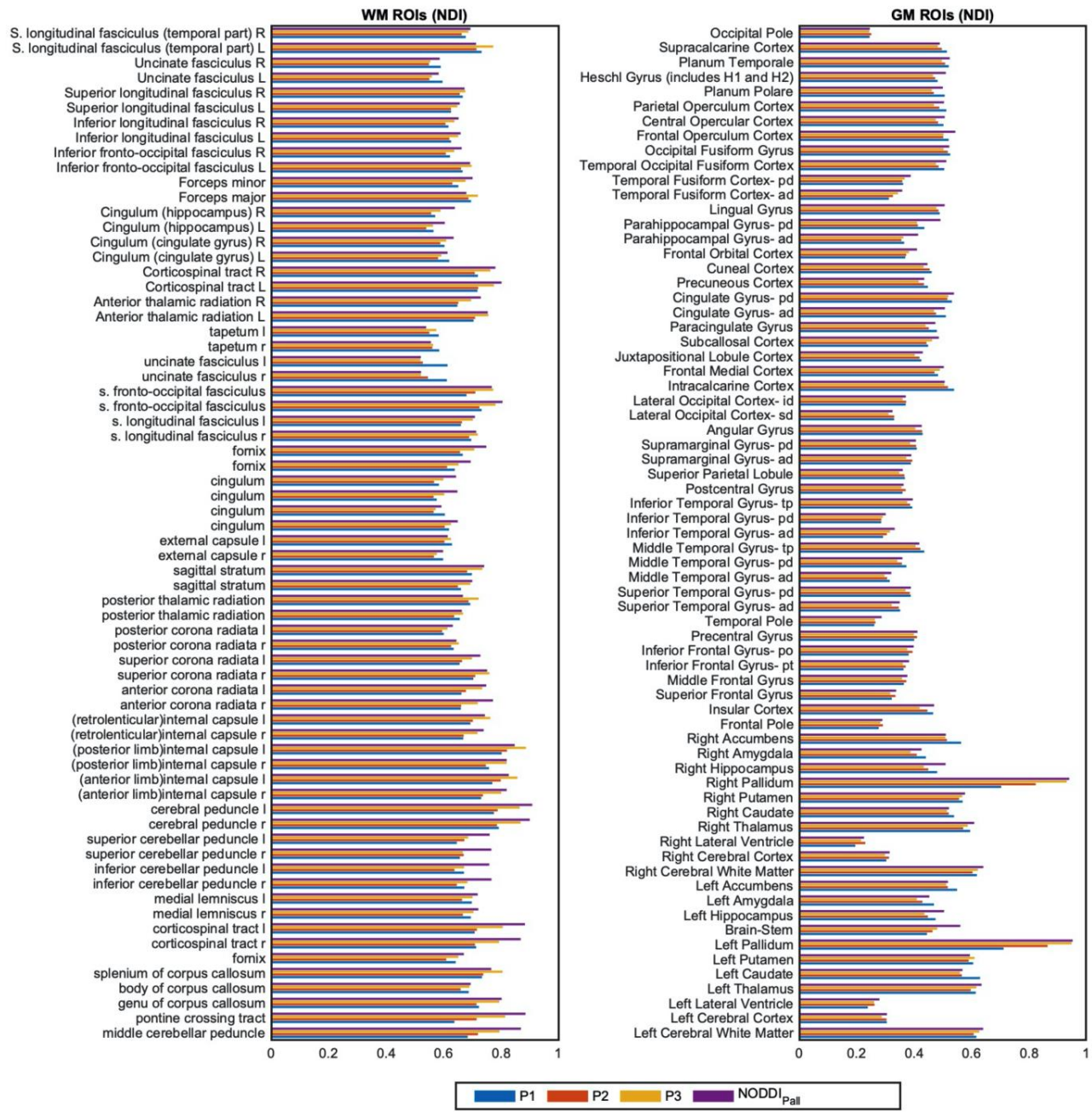

**Supplementary Figure S5. Comparison of ODI values in brain ROIs**

Comparison of orientation dispersion index (ODI) for DLpN single-shell protocols (P1, P2, P3) and multi-shell $NODDI_{Pall}$ for different white matter (WM) and grey matter (GM) ROIs using John Hopkins University (JHU) WM and WM tract atlas and Harvard-Oxford (HO) GM atlas.

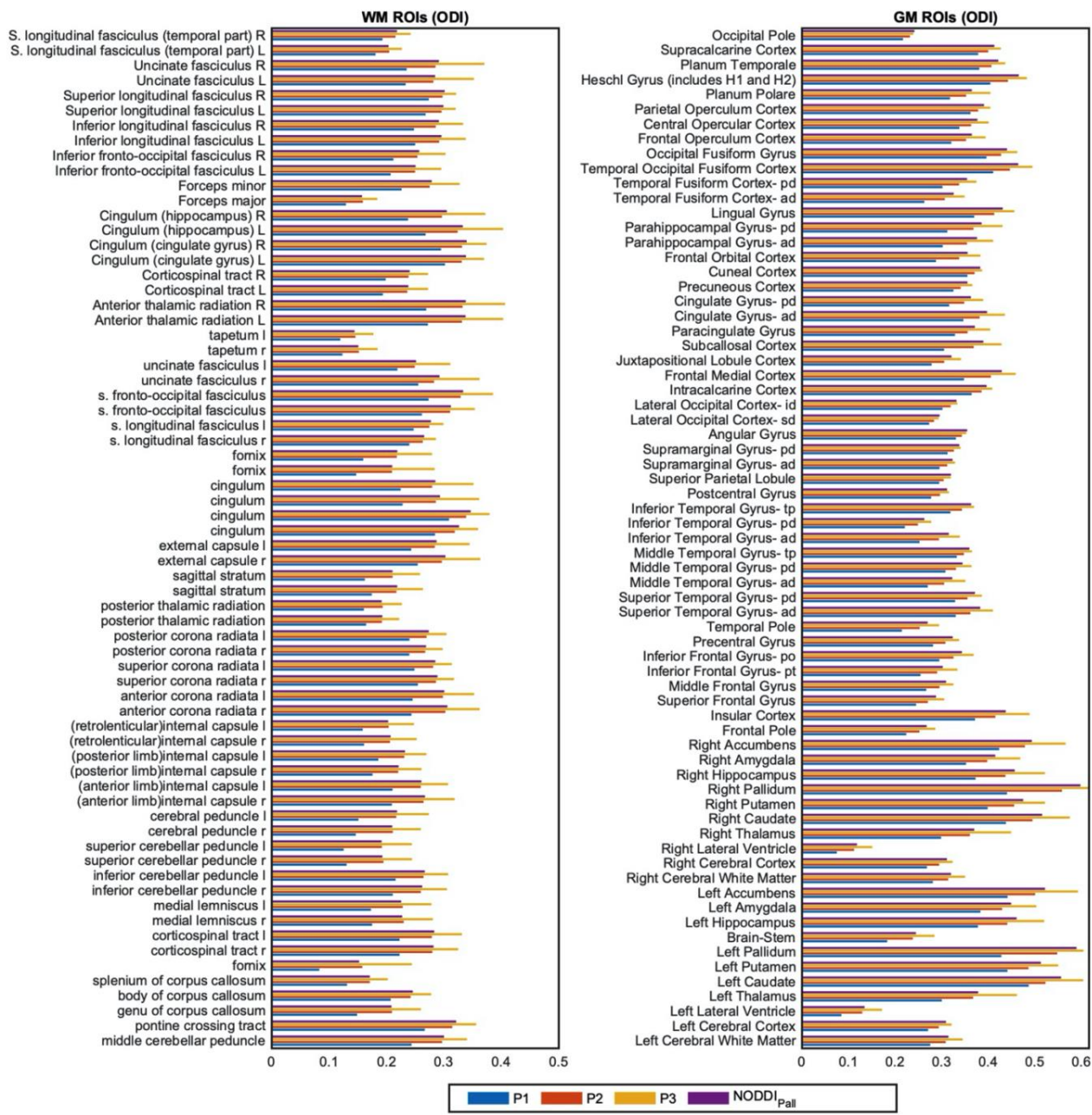

**Supplementary Figure S6. Correlation between DLpN derived NDI, ODI and DictNet derived $f_{ISO}$ with respective NODDI$_{Pall}$ derived maps**

Scatter plots showing significant linear correlations between DLpN derived NDI, ODI, and DictNet derived $f_{ISO}$ at different protocols (P1, P2, P3, P12, P13, P23 and Pall) with the pseudo ground-truth NODDI fitting with Pall (NODDI$_{Pall}$). Asteric symbols are the mean of all the ROIs from John Hopkins University White Matter (WM) and Harvard Oxford cortical and sub-cortical grey matter (GM) atlases. Gray indicating GM and blue indicating WM ROIs. Both single-shell and multi-shell protocol for DLpN showed very strong concordance with the pseudo ground-truth NODDI$_{Pall}$.

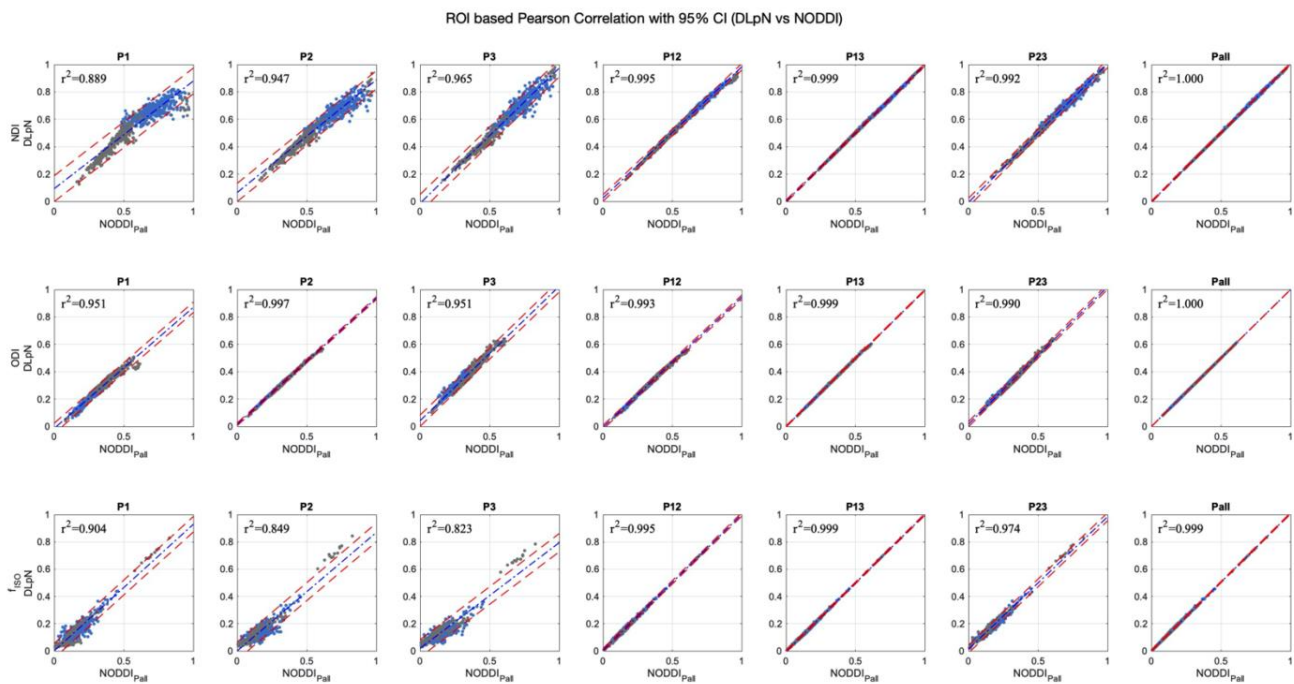

**Supplementary Figure S7. DTI based $f_{ISO}$ or FW used as prior for NDI reconstruction.**

For one of the HCP test subjects, we obtained multi-shell fitted DTI based FW (mFW) map using the dipy library and used it as prior to reconstruct NODDI based NDI. The first three images (from left to right) are the cases where multi-shell fitted FW (mFW) was used as prior for P1, P2 and P3 protocols respectively. The 4th one is our pseudo-Ground Truth NODDI$_{Pall}$ NDI and 5th one is the DLpN based reconstruction for the same subject with P2 i.e., protocol with b=2000 s/mm$^2$.

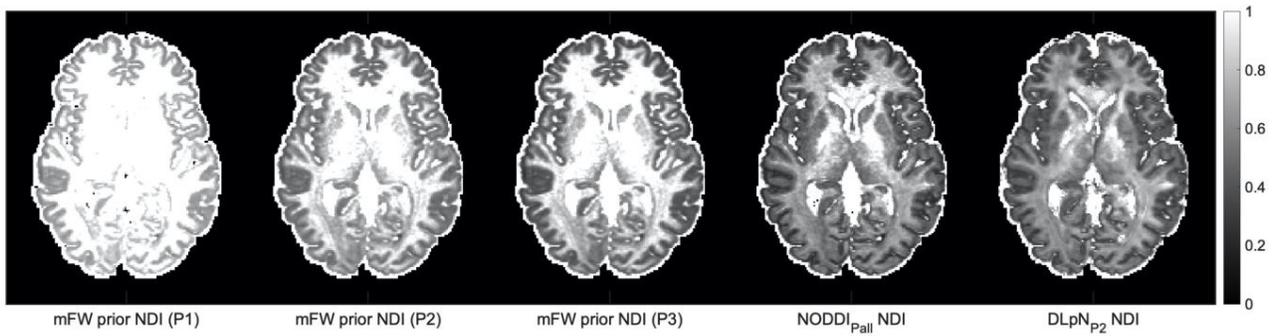

**Cerebrovascular Small Vessel Disease: Case Study for Single-Shell NDI**

*Protocol and Training Data Acquisition:* For one of our ongoing Cerebrovascular Small Vessel Disease (CSVD) we obtained written informed consent according to the institutional protocol and scanned recruited participants. For this study, one reflecting case of CSVD subject was chosen to test our approach. The dMRI scan was performed using 2D single-shot spin echo echo-planar imaging (SE-EPI) sequence (TR= 4300 ms; TE= 69.0 ms; ESP= 0.66 ms; 1.5 mm isotropic, 64 gradients per shell with b=1,000 and 2,000 s/mm$^2$ with 7 b=0 s/mm$^2$ reference images). In order to facilitate the training of our model, we scanned two volunteers with the same protocol as above along with an additional b=3000 s/mm$^2$ for the same 64 gradient directions.

*Application on the Test Subject:* We applied our method to the clinical case and compared with original 2-shell NODDI fitting and co-registered FLAIR in S7. DLpN P2 derived NDI shows that we can generate relevant NDI contrast in pathological cases. Also, the single shell reconstruction showed to minimize multi-shell fitting noise to a large extent maintaining relevant contrast in the pathology.

DLpN$_{P2}$ based NDI showed to reconstruct a clear NDI map than 2-shell fitted NODDI, as the prior DictNet f$_{ISO}$ accounts for neighbouring spatial information for each voxel under reconstruction. *It should be noted that NDI in the CSF region is undefined in the NODDI model, and the NODDI model characterizes NDI values to be 1 when encounters high f$_{ISO}$. In our reconstructed cases, NDI is close to 0 for CSF (high f$_{ISO}$) i.e. the regions where NDI is undefined.*

*Validation:* Using FLAIR and 2-shell reconstructed NODDI data, we were able to qualitatively validate the reconstruction obtained from DLpN. As our learner is trained on 3-shell NODDI data and tested on the middle shell, in order to quantitively validate the test data we would need 3-shell dMRI data which was unfortunately unavailable. But we provide quantitative validation on JHU WM and HO subcortical GM ROIs on 6 HCP test subjects where the 3-shell data was available for the test subjects in the main manuscript. (Figure 8, Supplementary Figures S3, S4, S5 and S6).

**Supplementary Figure S8. Comparison of NDI maps from DLpN$_{P2}$ and NODDI$_{Pall}$ in CSVD**

A) Coronal slices from a test subject with CSVD (characterized by high Fazekas score of 3, and lesion volume of 1.77 cm$^3$). The NDI maps computed with DLpN (single shell b=2000 s/mm$^2$) and original NODDI (two-shells b=1000, 2000 s/mm$^2$), and corresponding T2 FLAIR images showing lesions (indicated by red arrows). Lesions in both NDI maps are clearly visible. B) Glass brain 3D rendering of the lesions.

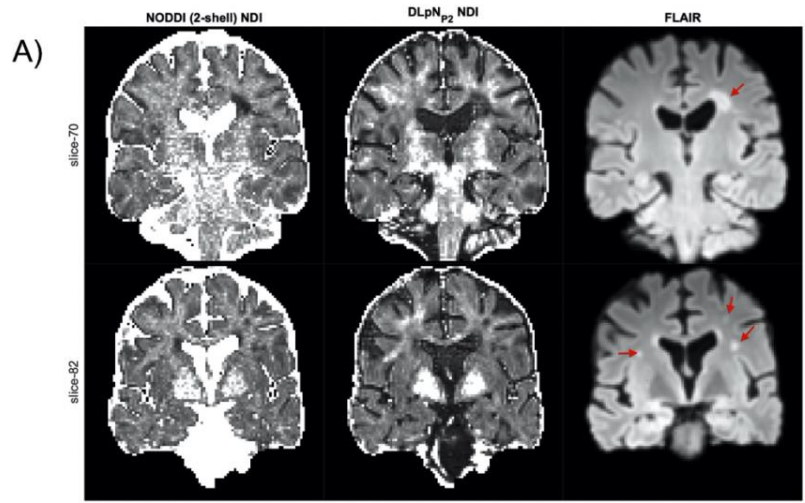

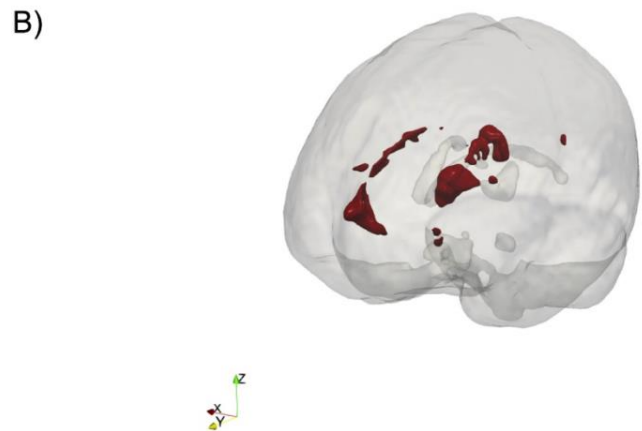